\documentclass[aps,prb,superscriptaddress,twocolumn]{revtex4-1}

\usepackage{graphicx}
\usepackage[dvipsnames]{xcolor}
\usepackage{amssymb}
\usepackage{physics}
\usepackage{bm}
\usepackage{cellspace,tabularx}
\usepackage{mathtools}
\usepackage[colorlinks=true,citecolor=blue,urlcolor=blue]{hyperref}
\usepackage{float}
\begin{document}

\title{
Localized Wannier function based tight-binding models for two-dimensional allotropes of bismuth
}

\date{\today}

\affiliation{Department of Materials Science and Engineering, Monash University, Clayton, VIC, 3168, Australia}
\affiliation{Chemical and Quantum Physics, School of Science, RMIT University, Melbourne, VIC, 3001, Australia}
\affiliation{ARC Centre for Future Low-Energy Electronics Technologies, Australia}

\author{Qile Li}
\affiliation{Department of Materials Science and Engineering, Monash University, Clayton, VIC, 3168, Australia}
\affiliation{ARC Centre for Future Low-Energy Electronics Technologies, Australia}

\author{Jackson~S.~Smith} 
\email{jackson.smith@rmit.edu.au}
\affiliation{Chemical and Quantum Physics, School of Science, RMIT University, Melbourne, VIC, 3001, Australia}
\affiliation{ARC Centre for Future Low-Energy Electronics Technologies, Australia}

\author{Yuefeng Yin}
\email{yuefeng.yin@monash.edu}
\affiliation{Department of Materials Science and Engineering, Monash University, Clayton, VIC, 3168, Australia}
\affiliation{ARC Centre for Future Low-Energy Electronics Technologies, Australia}

\author{Chutian Wang}
\affiliation{Department of Materials Science and Engineering, Monash University, Clayton, VIC, 3168, Australia}
\affiliation{ARC Centre for Future Low-Energy Electronics Technologies, Australia}

\author{Mykhailo~V.~Klymenko}
\affiliation{Chemical and Quantum Physics, School of Science, RMIT University, Melbourne, VIC, 3001, Australia}

\author{Jared~H.~Cole}
\email{jared.cole@rmit.edu.au}
\affiliation{Chemical and Quantum Physics, School of Science, RMIT University, Melbourne, VIC, 3001, Australia}
\affiliation{ARC Centre for Future Low-Energy Electronics Technologies, Australia}

\author{Nikhil~V.~Medhekar}
\email{nikhil.medhekar@monash.edu}
\affiliation{Department of Materials Science and Engineering, Monash University, Clayton, VIC, 3168, Australia}
\affiliation{ARC Centre for Future Low-Energy Electronics Technologies, Australia}

\begin{abstract}

With its monoelemental composition, various crystalline forms and an inherently strong spin-orbit coupling, bismuth has been regarded as an ideal prototype material to expand our understanding of topological electronic structures. 
In particular, two-dimensional bismuth thin films have attracted a growing interest due to potential applications in topological transistors and spintronics. 
This calls for an effective physical model to give an accurate interpretation of the novel topological phenomena shown by two-dimensional bismuth.
However, the conventional semi-empirical approach of adapting bulk bismuth hoppings fails to capture the topological features of two-dimensional bismuth allotropes because the electronic band topology is heavily influenced by crystalline symmetries as well as atom spacings. 
Here we provide a new parameterization using localized Wannier functions derived from the Bloch states in first-principles calculations. 
We construct new tight-binding models for three types of two-dimensional bismuth allotropes: a Bi (111) bilayer, bismuthene and a Bi(110) bilayer.
We demonstrate that our tight-binding models can successfully reproduce the band structures, symmetries and topological features of these two-dimensional allotropes. 
We anticipate that these models can be 
extended to other similar two-dimensional topological structures such as antimonene and arsenene. 
Moreover, these models can serve as a starting point for investigating the electron/spin transport and electromagnetic response in low-dimensional topological devices.
\end{abstract} 

\maketitle

\section{Introduction}

The discovery of spin-orbit coupling (SOC) induced topological phase transitions in electronic structure have led to a rapidly growing interest in the topological electronics and spintronics applications\cite{RevModPhys.82.3045,RevModPhys.88.021004}.  Due to an intrinsically strong SOC, bismuth and its compounds offer a rich playground for the development and application of topological band theories\cite{drozdov2014one}.
The electronic and topological structure of pure bismuth depends considerably on its crystal structure. Bulk bismuth is a topologically trivial semimetal despite a strong intrinsic SOC \cite{Liu1995a}. In contrast, when the bismuth lattice is confined to two dimensions, the resulting bismuth allotropes have been predicted to offer a rich spectrum of topologically non-trivial and distinct phases\cite{PhysRevLett.107.136805}.
For instance, the Bi (111) bilayer---a monolayer bismuth arranged in a buckled configuration---is predicted to have a quantum spin Hall (QSH) phase\cite{Murakami2006a}. On the other hand, bismuthene---a graphene-like planar layer of bismuth arranged in a honeycomb lattice---
 is reported to be a topological crystalline insulator\cite{munoz2016topological,Hsu2016a}.
Due to the diversity of non-trivial topological phases, two-dimensional bismuth allotropes have attracted extensive interest in recent years as a potential candidate for building novel topological electronic and spintronic devices\cite{Reis2017a,Schindler2018a,hsu2019topology}.

Recent experiments have demonstrated that manipulating the band topology of different 2D bismuth allotropes can lead to many exotic physical phenomena.
For example, it has been shown that the ultrathin Bi (111) films can be used to tune topological edge states when interfaced with other 2D materials\cite{Hirahara2011a,Wang2014a,Ma2015a}.
In a Bi (110) bilayer, elastic strains and external electric fields can significantly affect the stability of its topological phase due to its sensitivity to atomic buckling and charge doping\cite{Li2017a,Lu2015a}.
For planar bismuthene, recent experiments and calculations have demonstrated a controllable orbital-filtering QSH effect due to a selective bonding with silicon carbide substrate\cite{Reis2017a}.
Moreover, reports have also suggested the presence of exotic, higher-order topological hinge states in bismuth\cite{Schindler2018a}.
Despite the recent progress, a unified understanding of the correlation between the crystalline symmetries of 2D bismuth allotropes and their topological phases is still being developed.

Previous attempts to understand the influence of the crystal symmetries of 2D bismuth on electronic structure have largely been limited to semi-empirical tight-binding (TB) models of the bulk bismuth \cite{Liu1995a,PhysRevB.93.041301,NOURI2020126364}. 
However, as we illustrate later, these models cannot be used for 2D allotropes of bismuth for several reasons. First, the symmetry of two-dimensional bismuth is different from that of the bulk---the semi-empirical TB model can not faithfully reflect the symmetry reduction from bulk to a surface-like structure. 
Moreover, the semi-empirical model does not consider the relaxation of atomic positions in the two-dimensional layer relative to the bulk. A notable example is the planar honeycomb structure of bismuthene, which can not be directly related to the bulk bismuth symmetry.  
Consequently, these issues lead to a poor agreement between the band structure computed from the semi-empirical model and first principles density functional theory (DFT) \cite{Bieniek2017a}.

To gain physical insights into correlation between crystal symmetry and topological phases of 2D allotropes of bismuth, here we develop 
effective TB models of Bi (111) and Bi (110) bilayers as well as bismuthene using localized Wannier functions constructed from first-principles calculations\cite{Mostofi2008a}.
Wannier functions offer a natural choice of an orthonormal basis set due to the connection between the charge center of these functions and the Berry phase of Bloch states\cite{Zak1989a}.  These functions can be exponentially localized to either atom centers or interstitial sites, and are therefore similar to the atomic orbitals.
Furthermore, the Wannier basis set can be constructed without the need to fit free parameters to the band structures obtained from DFT or experiments.
Wannier functions thereby allow us to construct a model Hamiltonian for each allotrope with relatively few parameters and yet still provide an accurate description of their band structure and its relationship with the crystal symmetry. These Wannier function based Hamiltonians can be further employed to investigate charge and spin carrier transport in electronic devices based on low-dimensional topological materials.

This article is organized as follows. 
In Sec.~\ref{sec:semi_empirical_tight_binding} we discuss the limitations in the applicability of the semi-empirical TB parameters from bulk bismuth to a Bi (111) bilayer.
We then introduce a method for parameterising new TB models from the first principles calculations in Sec.~\ref{sec:wannier_tight_binding_models}.
First we outline our density-functional methods in Sec.~\ref{sec:density_functional_method}, then in Sec.~\ref{sec:wannier_method} we explain the construction of the TB models from localized Wannier functions, including constraints due to the symmetries of the crystallographic lattices.
Finally, in Sec.~\ref{sec:results} we discuss the results of our parameterizations for a Bi (111) bilayer (Sec.~\ref{sec:results_111_bilayer}), bismuthene (Sec.~\ref{sec:results_bismuthene}), and a Bi (110) bilayer (Sec.~\ref{sec:results_110_bilayer}).
We conclude in Sec.~\ref{sec:conclusion}.

\section{The applicability of tight binding parameters from bulk to a bilayer}
\label{sec:semi_empirical_tight_binding}

\begin{figure}
  \includegraphics{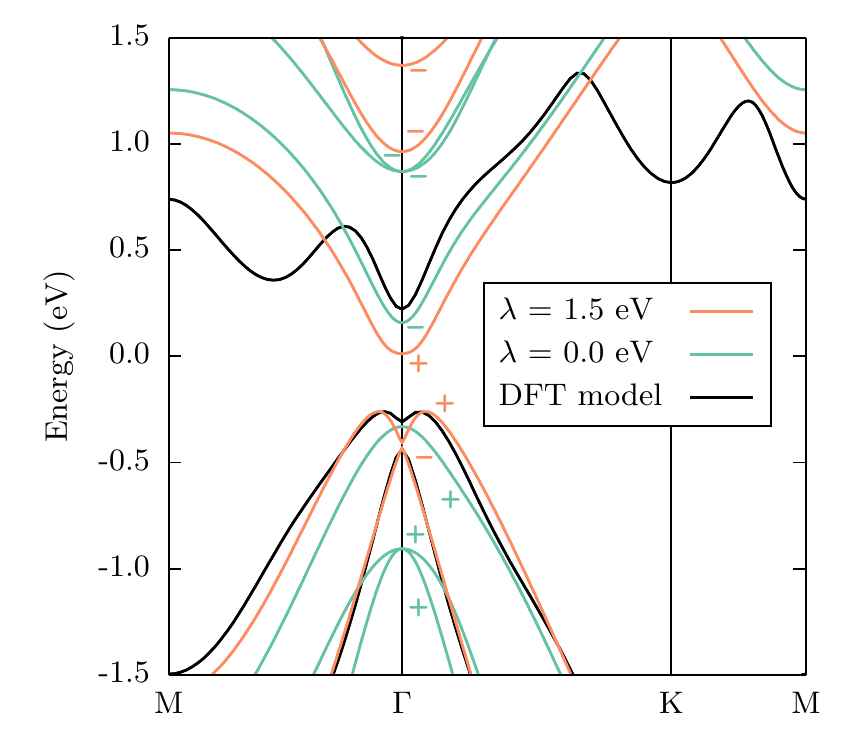}
  \caption{Band structures of a Bi (111) bilayer computed from TB parameters of bulk bismuth~\cite{Liu1995a} with (orange lines) and without (green lines) SOC. The parity of each eigenstate at the $\Gamma$ is labelled as either even ($+$) or odd ($-$). The band structure predicted by DFT (black lines) has been lowered by 0.25~eV to align 	 the valence bands of the two models.}
  \label{fig:111_bilayer}
\end{figure}



\begin{figure}
  \includegraphics[width=1.0\columnwidth]{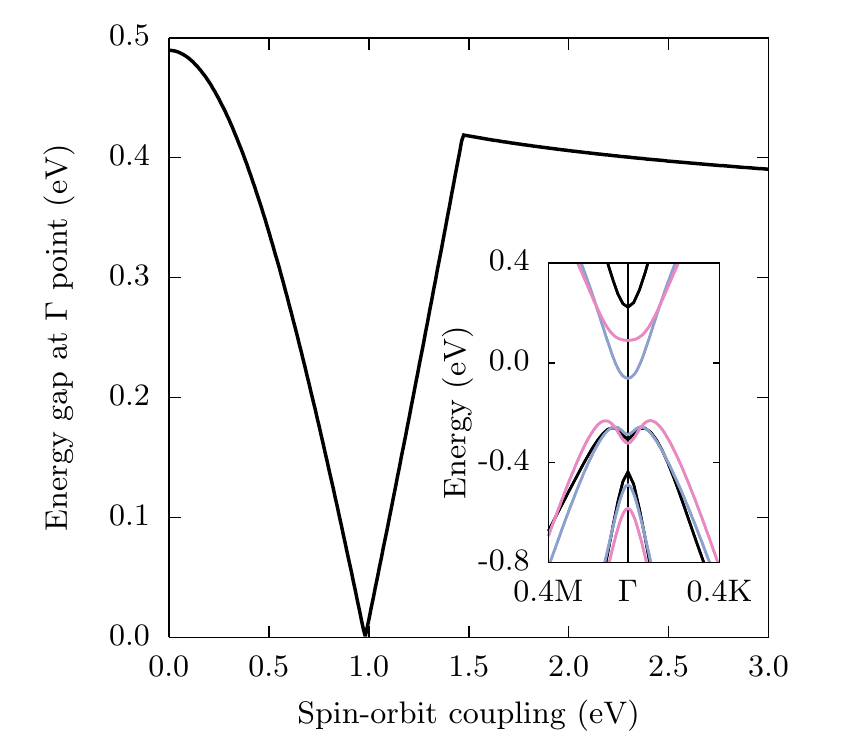}
  \caption{Energy gap between the valence and conduction bands at the $\Gamma$ point computed from TB parameters of bulk bismuth~\cite{Liu1995a} as a function of SOC. The inset shows band structures of a Bi (111) bilayer computed from the same parameters with SOCs of 1.25~eV (blue lines) and 1.75~eV (pink lines). The band structure of DFT is also shown in the inset (black lines).}
  \label{fig:111_bilayer-3}
\end{figure}

The electronic properties of Bi (111) bilayers\cite{Murakami2006a,Bieniek2017a} are typically modeled using semi-empirical TB parameters derived for bulk bismuth\cite{Liu1995a}.
For example, this strategy was adopted in Ref. \citenum{Murakami2006a} and \citenum{Bieniek2017a}, 
by using the hopping parameters of bulk bismuth to bismuth bilayer, 
with the SOC strength ($\lambda$) increased from its bulk value to better fit the energy splittings between the bands of a bilayer when compared to DFT calculations.
In the following we will investigate the validity of this approach, as outlined in Ref. \citenum{Murakami2006a} and \citenum{Bieniek2017a}, 
and determine if using bulk tight-binding parameters to two dimensional structure can faithfully reproduce all critical electronic features of the topologically nontrivial Bi (111) bilayer.

We use TB parameters for bulk bismuth to calculate the band structure of a Bi (111) bilayer but truncate these parameters by retaining only the hoppings inside the bilayer\cite{Murakami2006a}.
All calculations were performed using NanoNet, an extendable Python framework for electronic structure computations based on the tight-binding method \cite{KLYMENKO2021107676}.
The resulting band structure is shown in the colored lines of Fig.~\ref{fig:111_bilayer}. The SOC strength is taken from the bulk value (1.5 eV).
In this figure we have also plotted a band structure for the bilayer computed from DFT (black lines) for reference (see Sec.~\ref{sec:density_functional_method} for method).
There is good agreement between the valence bands of TB and DFT for a SOC strength of 1.5 eV but the agreement between the conduction bands is poor.
The lowest conduction band of the TB model has energy minima at points $\Gamma$ and M whereas DFT predicts two additional minima, one at K and another along the high symmetry segment $\mathrm{M} \to \Gamma$.

Another property of interest is band inversion, \textit{i.e.} the exchange of a crystal's electronic properties (\textit{e.g.} parity) between its conduction and valence bands.
The parity of each eigenstate at the $\Gamma$ point in Fig.~\ref{fig:111_bilayer} is labeled as either even ($+$) or odd ($-$).
The parities of the bands shown in the figure are even for the valence bands and odd for the conduction bands with no SOC ($\lambda = 0.0~\mathrm{eV}$).
When we use the bulk SOC value of 1.5 eV,  the parity of the second highest valence band and lowest conduction band are exchanged.
This is in good agreement with a previous study that assumes a Bi (111) film retains the bulk lattice parameter at about 20--50 nm of film thickness \cite{PhysRevB.83.121310}. 
However, several reports have shown that when bismuth films are reduced to only one bilayer thickness, the change of parities occurs between the highest valence band and lowest conduction band, contrary to the data presented in Fig.~\ref{fig:111_bilayer}.\cite{Li2014a,PhysRevLett.107.166801} This contrasting behavior has been associated with the change of the lattice parameter from its bulk value.\cite{Li2014a,PhysRevLett.107.166801}

SOC also significantly affects the band gap in a Bi (111) bilayer.  
Fig.~\ref{fig:111_bilayer-3} shows the variation of the band gap at $\Gamma$ point between the highest valence band and the lowest conduction band as a function of the SOC strength.
The maximum energy gap that can be achieved at the $\Gamma$ point by increasing the strength of the SOC is approximately 0.4~eV.
This gap is smaller than the gap computed from DFT (see inset of Fig.~\ref{fig:111_bilayer-3}) which typically underestimates band gaps compared to their experimental values\cite{Remediakis1999a}.
We have found that the band splitting agrees with DFT results if we increase the SOC strength in the TB model to 1.8 eV, as suggested by Ref.~\citenum{Bieniek2017a}. However, the band shape and the exchange of parities still do not agree with \textit{ab initio} calculations and experimental observations \cite{Li2014a,PhysRevLett.109.016801}. 
Based on the results above, we can conclude that an effective tight-binding model of two-dimensional bismuth layered materials can not be achieved by using bulk hopping parameters, even by tuning the strength of SOC. 
Consequently, a new model is needed for 2D allotropes of bismuth that adapts to the structural changes of the material when reduced from the bulk to a thin film.

\begin{figure*}[htbp]
  \includegraphics[width=2.00\columnwidth]{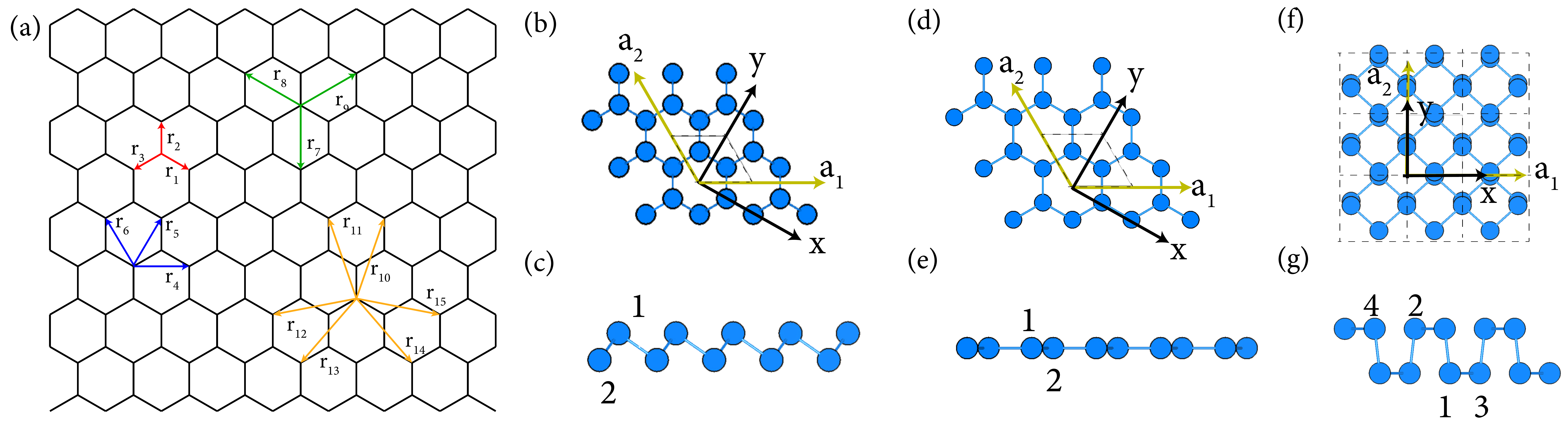}
  \caption{ (a) Top-down view of first-nearest neighbor (read arrows), second-nearest neighbor (blue arrows), third-nearest neighbor (green arrows), and fourth-nearest neighbors (orange arrows) hopping vectors of bismuth allotropes.(b-g)Schematics of the two-dimensional allotropes of bismuth, with atoms (blue spheres) and primitive cells (solid black lines). (b) Top down view and (c) side view of a Bi (111) bilayer. (d) Top down view and (e) side view of bismuthene. (f) Top down view and (g) side of a Bi (110) bilayer. The unit cells of these bismuth allotropes are shown in  (b), (d) and (f) in dashed lines. In subfigures (c), (e), and (g), each atom in the primitive cell is labeled with an indice ranging from 1 to 4. The labels a\textsubscript{1}, a\textsubscript{2} and x, y denote the lattice vector and cartesian axis respectively.}
  \label{atom_schematics}
\end{figure*}

\section{Tight binding models from Wannier functions}
\label{sec:wannier_tight_binding_models}

\subsection{Density-functional method}
\label{sec:density_functional_method}

We first obtained the electronic structure of the 2D allotropes of bismuth through DFT as implemented in the Vienna Ab initio Simulation Package (VASP)\cite{Kresse1996a,Kresse1996b}.
Exchange and correlation effects were captured by the generalized gradient approximation and the Perdew-Burke-Ernzerhoff functional\cite{Perdew1996a}.
The atomic structures of the allotropes were first optimized prior to a self-consistent convergence of the electronic structure using a $\Gamma$-centered  $21 \times 21 \times 1$ $k$-point grid.
The energy cutoff for the plane-wave basis used was 600~eV. Wavefunctions obtained using DFT were utilized to construct  TB models based on Wannier functions using the code Wannier90\cite{Mostofi2014a}, as discussed in the next section.

\subsection{Construction of Wannier tight binding models and symmetry constraints}
\label{sec:wannier_method}

For periodic systems with translational symmetry, the one-particle states can be expressed by the Bloch state $\ket{\psi_{n\bm{k}}}$ with band index $n$ and crystal momentum $\bm{k}$.
When deriving parameters for the hopping of an electron from one orbital in a crystal to another it is more convenient to consider localized orbitals rather than Bloch states because the latter are delocalised.
Wannier functions are one such choice of localised orbitals and they can be constructed by the inverse Fourier transform of a Bloch state.
In this way we can obtain a real-space Wannier TB Hamiltonian by a discrete Fourier transformation\cite{Marzari2012a}:
\begin{equation} \label{WTB-R}
\bra{\bm{0}} \hat{H} \ket{\bm{R}} = \frac{1}{N} \sum_{\bm{q}} e^{-i\bm{q} \cdot \bm{R}} U_{\bm{q}}^{\dagger} E_{\bm{q}} U_{\bm{q}}
\end{equation}
where $\bm{R}$ is a real-space lattice vector, $\bm{0}$ is the real-space lattice vector that defines the home cell [\textit{i.e.} $\bm{R} = (0, 0, 0)$], $N$ is the number of points in the $k$-point grid and $E_{\bm{q}}$ are eigenvalues from DFT.
$U_{\bm{q}}$ is a unitary transformation that takes the Bloch state at point $\bm{q}$ in $k$-space to a rotated Bloch state in the Wannier gauge\cite{Marzari2012a}, \textit{i.e.} 

\begin{equation}\label{TR-U}
\ket{\tilde{\psi}_{n\bm{k}}} = \sum_{m} U_{mn}^{(\bm{k})} \ket{\psi_{m\bm{k}}}
\end{equation}
where $n$ and $m$ are band indices.
The corresponding reciprocal space TB Hamiltonian $H_{\alpha\beta}(\bm{k})$ can then be expressed as a Fourier transformation from the real space Hamiltonian $H_{\alpha\beta}(\bm{R})$ to $k$ space:
\begin{equation} \label{WTB-K}
H_{\alpha\beta}(\bm{k})=\sum_{\bm{R}} e^{i\bm{k}\cdot(\bm{R}+\bm{\tau_\beta}-\bm{\tau_\alpha})} H_{\alpha\beta}(\bm{R})
\end{equation}
where the subscript $\alpha\beta$ denotes a Hamiltonian matrix element  that corresponds to hopping from orbital $\alpha$ at $\bm{\tau}_{\alpha}$ in the home cell to orbital $\beta$ at $\bm{\tau_\beta}$ within a cell at $\bm{R}$. $H_{\alpha\beta}(\bm{R})$ can be further expanded as:
\begin{equation} \label{WTB-R-T}
H_{\alpha\beta}(\bm{R})= t_{\alpha\beta} (\bm{R}-\bm{0}) = \bra{\bm{0}+\tau_{\alpha}} H \ket{\bm{R}+\tau_{\beta}}
\end{equation}
where $t_{\alpha\beta} (\bm{R}-\bm{0})$ represents the hopping parameter between neighboring atomic orbitals $\alpha$ in the home cell and $\beta$ in the cell $\bm{R}$ extracted from the Wannier TB Hamiltonian. For simplicity, we have used simple numerical subscript $t_{n}$ in later presentation of our TB models. A detailed discussion of the extraction process and the role of each hopping parameter can be found in the Appendix.

Next we discuss the symmetry constraints on this Hamiltonian.
The representations generated from the method implemented in Wannier90\cite{Gresch2018a} often contain small numerical errors that can break the symmetries of the crystal's energy bands.
We correct these errors and restore the corresponding constraints on the crystal's symmetry.
Our Hamiltonians satisfy the following symmetry constraints in $k$-space and real-space respectively\cite{Gresch2018a}:
\begin{align}
H({\bm{k}}) &= D(g)H(g^{-1}{\bm{k}})D^{-1}(g) &&g \in G \label{eq5} \\
H_{ij}(\bm{R^{\prime}}) &= \sum_{\alpha, \beta} D_{i\alpha}(g) H_{\alpha\beta}(\bm{R}) D^{-1}_{\beta j}(g) &&g \in G \label{eq6}
\end{align}
$D(g)$ is the matrix representation of the symmetry operation $g$, which is an element of symmetry group $G$.
The matrix representations used for the 2D allotropes of bismuth ($D_{3d}$ for the Bi (111) bilayer, $D_{6h}$ for bismuthene and $D_{2h}$ for the Bi (110) bilayer) are given in the Appendix.
The subscripts $\alpha$ and $\beta$ are the orbital indices before the symmetry operation and $i$ and $j$ are the corresponding indices after this operation.
$\bm{R^{\prime}}$ is a lattice vector that defines the cell's position after the symmetry operation.

To ensure our Hamiltonian matrix satisfies each symmetry constraint for a particular allotrope, we take a group average over all Hamiltonians transformed by the symmetry operations in $G$ through Eq.~\ref{eq6}.
This yields a symmetrized Hamiltonian\cite{Gresch2018a}:
\begin{equation} \label{eq9}
\tilde{H}_{\alpha\beta}(\bm{R}) = \frac{1}{l} \sum_G D_{i \alpha}^{-1}(g) H_{i j}(\bm{R^{\prime}}) D_{\beta j}(g),
\end{equation}
where $l$ is the number of elements in the symmetry group.

\section{Results}
\label{sec:results}

To constrain our basis functions to crystal symmetries and fix Wannier orbitals to atomic positions, we construct our TB Hamiltonian from a spinless case without performing maximal localization on Wannier Functions. We have also treated SOC independently, evaluating it in the basis of atomic orbitals and then fitting the SOC parameters by comparing to the bands of DFT. In the following subsections we will discuss the symmetry of the crystal structures, properties of the basis functions, and construction of the TB models for each allotrope.
 
It should also be noted that we will emphasize the necessity of determining the symmetry properties of electronic bands in constructing the TB model for each allotrope. The SOC effects are also treated differently for each bismuth allotrope considering the changes in crystalline symmetries. 
We calculate the matrix representations of all the symmetry operations for eigenstates at high symmetry points.
Finally we confirm the accuracy of our TB models by calculating the basis functions for these irreducible representations using the projection operator method\cite{Dresselhaus2008a} and check against orbital characters obtained using DFT calculations for consistency.

\subsection{Bi (111) bilayer}
\label{sec:results_111_bilayer}

\begin{figure*}[htbp]
  \includegraphics{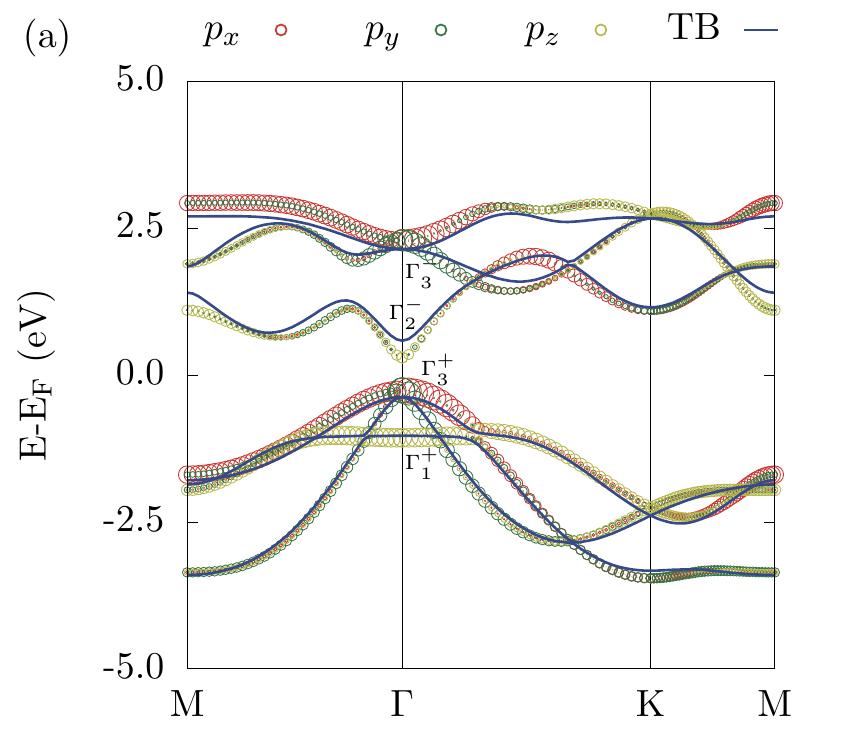}
  \includegraphics{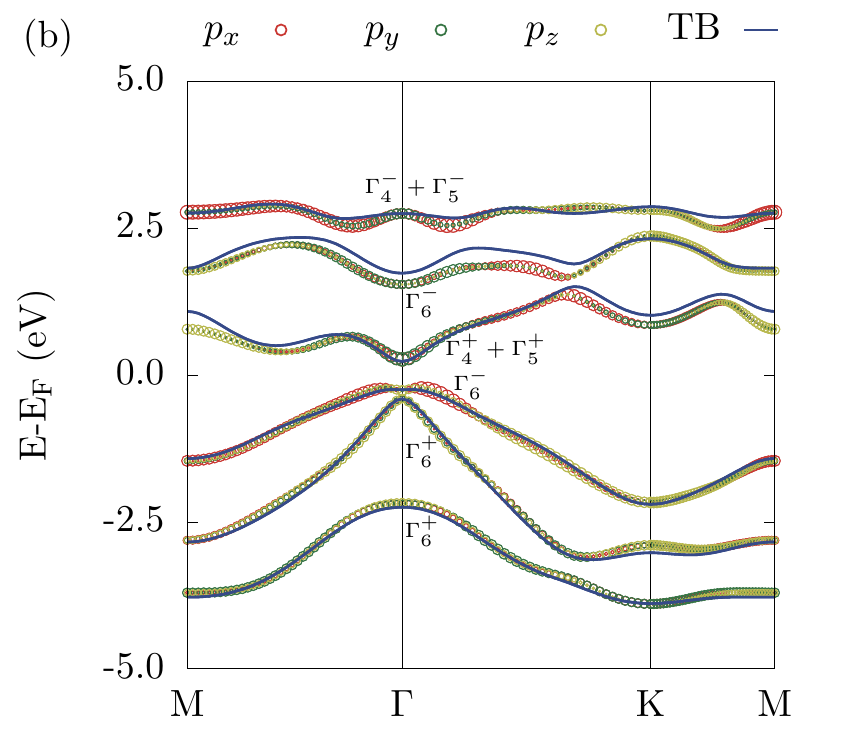}
  \caption{Bilayer bismuth (111) band structure (a) without and (b) with SOC by including neighboring interactions up to fourth NNs. Energy bands from our TB model are shown as solid dark blue lines. Energy bands from DFT are plotted as circles, with $p_z$ in yellow, $p_x$ in red and $p_y$ in green. The weights of orbitals are represented by the size of circles. }
  \label{figbi111band}
\end{figure*}

The Bi (111) bilayer has a quasi-2D honeycomb structure characterized by an out-of-plane buckling with an intra-layer spacing of approximately 0.87~\AA.
There are two inequivalent atoms in the primitive cell as shown in Fig. \ref{atom_schematics}.
The crystal structure of the Bi (111) bilayer has a symmorphic space group symmetry of $P\overline{3}m1$ ($D_{3d}^{3}$, SG164).
The symmetry generators for its point subgroup include the identity ($E$), inversion ($P$), threefold rotation along the $z$ direction ($C_{3z}$), and three twofold rotations ($C_{2}$).
From the orbital characters obtained from our DFT calculations, we find the $s$ orbitals form an isolated set of bands well below the Fermi level.
Closer to the Fermi level, $p$ orbitals form two sets of isolated bands: three valence bands (VBs) and three conduction bands (CBs).
We therefore choose to include only $p$ orbitals in our TB model.

The number of nearest neighbor (NN) hoppings to include is a critical parameter for a TB model. 
To investigate the relationship between neighbor interactions and the agreement of our model with DFT, we calculate the average energy difference between the two models along a path of high symmetry in the first Brillouin zone (FBZ).
We find that a Hamiltonian with second NN interactions can well reproduce the topological properties of the Bi (111) bilayer, with good agreement to the occupied states.
However, the inclusion of up to fourth NNs is needed to recover the features of unoccupied bands (see Fig. \ref{figbi111band} (b)), but at the cost of requiring more hopping parameters. 
In the following we will focus more on the details of building a TB model based on second NN interactions. The model considering fourth NNs can be simply constructed in a similar way by incorporating more neighbor interactions.

Next we discuss the steps involved in constructing our TB model in more detail.
We start from a spinless Hamiltonian written in a basis of $p$-like orbitals.
The basis set is \{$p_{z}^{(1)}$, $p_{x}^{(1)}$, $p_{y}^{(1)}$, $p_{z}^{(2)}$, $p_{x}^{(2)}$, $p_{y}^{(2)}$\} where (1) and (2) denote each of the two atoms in the primitive cell, as labelled in Fig.~\ref{atom_schematics} (c).
We use a convention where the real space Hamiltonian is dependent on the lattice vectors $\bm{R}$ and sublattice vectors $\bm{\tau}_{\alpha}$ and $\bm{\tau}_{\beta}$ that correspond to particular atomic sites.
We construct hopping parameters within a group of NN interactions by applying symmetry operations to the Hamiltonian (which in turn depends on $\bm{R}$, $\bm{\tau_{\alpha}}$, and $\bm{\tau_{\beta}}$).
It is thereby possible to generate all hopping parameters within a given set of NN interactions [\textit{i.e.} first, second, third, or fourth, as shown in Fig.~\ref{atom_schematics} (a)] from the interactions between one pair of neigbor atoms in the set. We denote $\bm{r}_{1} $-- $\bm{r}_{3} $ as nearest neighbor hopping vectors, $\bm{r}_{4} $-- $\bm{r}_{6} $ as second-nearest neighbor hopping vectors, $\bm{r}_{7} $-- $\bm{r}_{9} $ as third-nearest neighbor hopping vectors, and $\bm{r}_{10} $-- $\bm{r}_{15} $ as fourth-nearest neighbor hopping vectors
The symmetry operations not only generate hopping parameters but also enforce the symmetry constraints that are defined in Eqs.~\ref{eq5} and \ref{eq6}.

The $k$-space TB Hamiltonian can then be constructed from the real-space Hamiltonian using Equation \ref{WTB-K}.
The $k$-space tight-binding Hamiltonian consists of sub-blocks related to hoppings between two sub-lattice atoms as shown in Equation \ref{Hk-form_GEN}: 
\begin{equation} \label{Hk-form_GEN}
H(\bm{k})=
\begin{pmatrix}
H^{11}(\bm{k}) & H^{12}(\bm{k}) \\
H^{21}(\bm{k})& H^{22}(\bm{k})\\
\end{pmatrix}
\end{equation}
Here we use the superscript to denote interactions between atoms and subscript for interactions between orbitals in Hamiltonian $H$.
The diagonal blocks $H^{11}(\bm{k}) $ and $H^{22}(\bm{k}) $ are on-site energies of the two bismuth atoms (Bi1 and Bi2) in the home cell, while $H^{12}(\bm{k}) $ and $H^{21}(\bm{k})$ are hopping matrix between the two bismuth atoms. The matrix elements in every $H^{ij}(\bm{k}) $ are therefore denoted as $H_{ij}(\bm{k}) $, which represents the interactions between orbitals. The indices 1, 2, ..., 6 correspond to the obtials \{$p_{zBi1}$, $p_{xBi1}$, $p_{yBi1}$, $p_{zBi2}$, $p_{xBi2}$, $p_{yBi2}$\}. We have dropped $k$ dependence in the sub-blocks for brevity:

\begin{equation}
H^{11}(\bm{k}) =
\begin{pmatrix}
H_{11}& H_{12}& H_{13} \\
& H_{22} & H_{23} \\
h.c. & & H_{33} \\
\end{pmatrix}
\end{equation}

\begin{equation}
H^{12}(\bm{k}) =
\begin{pmatrix}
H_{14} & H_{15} &H_{16}\\
H_{15}& H_{25} &H_{26}\\
H_{16} &H_{26} &H_{36}\\
\end{pmatrix}
\end{equation}

$H^{22}(\bm{k}) $ and $H^{21}(\bm{k})$ can be obtained by applying an inversion operation to $H^{11}(\bm{k}) $ and Hermit adjoint to $H^{12}(\bm{k})$ respectively:

\begin{equation}\label{inv}
H^{22}(\bm{k}) = H^{11}(\bm{-k})
\end{equation}
\begin{equation}\label{simplify1}
H^{12}(\bm{k}) = H^{21}(\bm{k})^{\dagger}
\end{equation}

Constrained by the inversion symmetry and Hermitian condition, $H^{12}(\bm{k})$ is symmetric. Hence, we only need to solve for $H^{11}(\bm{k}) $ and $H^{12}(\bm{k})$.
$H_{ii}(\bm{k})$ represents the self-interaction energy of an orbital with itself, which can be expressed as:

\begin{equation}\label{diag_1}
H_{ii}(\bm{k})=\epsilon_i+2h_{ii}^{(1)}[\cos(\bm{k}\cdot\bm{r_4})+\cos(\bm{k}\cdot\bm{r_6})]+2h_{ii}^{(2)}\cos(\bm{k}\cdot\bm{r_5})
\end{equation} 
where $\epsilon_1, \epsilon_2, \epsilon_3=t_1, t_2, t_2$; $h_{11}^{(1)}, h_{22}^{(1)}, h_{33}^{(1)} = t_7, t_{11}, t_{15}$; $h_{11}^{(2)}, h_{22}^{(2)}, h_{33}^{(2)}=t_7, (\frac{3}{2}t_{15}-\frac{1}{2}t_{11}), (\frac{3}{2}t_{11}-\frac{1}{2}t_{15})$. Here $t_{i}$ denote real space hopping parameters as defined in Eq. \ref{WTB-R-T}. 

For the remaining off-diagonal elements $H^{11}(\bm{k})$, they are given by:

\begin{equation}
\begin{aligned}\label{offdiag_12}
H_{12}(\bm{k})=2ih_{12}^{(1)}\sin(\bm{k}\cdot\bm{r_5})&+h_{12}^{(2)}(e^{i\bm{k}\cdot\bm{r_4}}+e^{-i\bm{k}\cdot\bm{r_6}})\\
&+h_{12}^{(3)}(e^{-i\bm{k}\cdot\bm{r_4}}+e^{i\bm{k}\cdot\bm{r_6}})
\end{aligned}
\end{equation} 

\begin{equation}\label{offdiag_13}
\begin{aligned}
H_{13}(\bm{k})=2ih_{13}^{(1)}\sin(\bm{k}\cdot\bm{r_5})&+h_{13}^{(2)}(e^{i\bm{k}\cdot\bm{r_4}}-e^{-i\bm{k}\cdot\bm{r_6}}) \\
&+h_{13}^{(3)}(e^{-i\bm{k}\cdot\bm{r_4}}-e^{i\bm{k}\cdot\bm{r_6}})
\end{aligned}
\end{equation} 

\begin{equation}\label{offdiag_23}
\begin{aligned}
H_{23}(\bm{k})=2ih_{23}^{(1)}\sin(\bm{k}\cdot\bm{r_5})&+h_{23}^{(2)}(e^{i\bm{k}\cdot\bm{r_4}}-e^{-i\bm{k}\cdot\bm{r_6}})\\
&+h_{23}^{(3)}(e^{-i\bm{k}\cdot\bm{r_4}}-e^{i\bm{k}\cdot\bm{r_6}})
\end{aligned}
\end{equation} 

where the constants $h_{ij}^{(m)}$ are related to hopping parameters by: $h_{12}^{(1)}=(-\frac{1}{2}t_8+\frac{\sqrt{3}}{2}t_9)$, $h_{12}^{(2)}=t_8$, $h_{12}^{(3)}=(-\frac{1}{2}t_8-\frac{\sqrt{3}}{2}t_9)$, $h_{13}^{(1)}=(\frac{1}{2}t_8+\frac{\sqrt{3}}{2}t_9)$, $h_{13}^{(2)}=t_9$, $h_{13}^{(3)}=(-\frac{\sqrt{3}}{2}t_8+\frac{1}{2}t_9)$, $h_{23}^{(1)}=(\frac{1}{2}t_{14}-\frac{1}{2}t_{12})$, $h_{23}^{(2)} = t_{12}$, and $h_{23}^{(3)} = t_{14}$ respectively. \\

For off-diagonal hopping sub-block $H^{12}(\bm{k})$, the matrix elements can be similarly expressed as:

\begin{equation}\label{offdiag_14}
H_{14}(\bm{k}) = h_{14}^{(1)}( e^{i\bm{k}\cdot\bm{r_1}}+e^{i\bm{k}\cdot\bm{r_2}}+e^{i\bm{k}\cdot\bm{r_3}})
\end{equation}

\begin{equation}\label{offdiag_15}
H_{15}(\bm{k}) = h_{15}^{(1)}(e^{i\bm{k}\cdot\bm{r_1}}-\frac{1}{2}e^{i\bm{k}\cdot\bm{r_2}}-\frac{1}{2}e^{i\bm{k}\cdot\bm{r_3}})
\end{equation}

\begin{equation}\label{offdiag_16}
H_{16}(\bm{k}) = h_{16}^{(1)}(e^{i\bm{k}\cdot\bm{r_2}}-e^{i\bm{k}\cdot\bm{r_3}})
\end{equation}

\begin{equation}\label{offdiag_25}
H_{25}(\bm{k}) = h_{25}^{(1)}e^{i\bm{k}\cdot\bm{r_1}}+h_{25}^{(2)}e^{i\bm{k}\cdot\bm{r_2}}+h_{25}^{(2)}e^{i\bm{k}\cdot\bm{r_3}}
\end{equation}

\begin{equation}\label{offdiag_26}
H_{26}(\bm{k}) = h_{26}^{(1)}e^{i\bm{k}\cdot\bm{r_2}}-h_{26}^{(1)}e^{i\bm{k}\cdot\bm{r_2}}
\end{equation}

\begin{equation}\label{offdiag_36}
H_{36}(\bm{k}) = h_{36}^{(1)}e^{i\bm{k}\cdot\bm{r_1}}+h_{36}^{(2)}e^{i\bm{k}\cdot\bm{r_2}}+h_{36}^{(2)}e^{i\bm{k}\cdot\bm{r_3}}
\end{equation}

Again, $h_{14}^{(1)}=t_3$, $h_{15}^{(1)}=t_4$, $h_{25}^{(1)}=t_5$, $h_{25}^{(2)}=(\frac{1}{4}t_5 + \frac{3}{4} t_6)$,$h_{26}^{(1)}=(-\frac{\sqrt{3}}{4}t_5+\frac{\sqrt{3}}{4}t_6)$, $h_{36}^{(1)}=t_6$ and $h_{36}^{(2)}=(\frac{3}{4}t_5 + \frac{1}{4}t_6)$. 

Following this procedure, we obtain the spinless TB Hamiltonian with 12 independent parameters ($t_1$ -- $t_{15}$, except $t_{10}$, $t_{13}$ and $t_{14}$)  for Bi (111) bilayer with up to the next-nearest neighbor interactions as summarized in Table \ref{TB_bilayer_111}.
If one desires a better description of the conduction bands, third and fourth NN interactions can be included. As shown in Table \ref{TB_bilayer_111}, a TB model involving up to four NN interactions included needs 10 additional independent hopping parameters ($t_{16}$ -- $t_{27}$, except $t_{18}$ and $t_{20}$). 
In the Appendix, we discuss the symmetry constraints imposed on the real space Hamiltonian and give the explicit forms of the real space Hamiltonian up to fourth NN intaractions. 

\begin{table}[h]
\caption{Hopping parameters for Bi (111) bilayer, up to and including fourth NN. Parameters with an asterisk superscript are related with other parameters in the table and they can be calculated by Eq. 36-38 in the Appendix. }
\label{TB_bilayer_111}
\begin{tabular*}{\columnwidth}{@{\extracolsep{\fill}} c r c r}
\hline
\hline
Parameter & Energy (eV) & Parameter & Energy (eV) \\
\hline
$t_{1}$ & -2.762 & $t_{9}$ & 0.160 \\
$t_{2}$ & -2.461 & $t_{10}$ & -0.226* \\
$t_{3}$ & 0.292 & $t_{11}$ & 0.211 \\
$t_{4}$ & -1.078 & $t_{12}$ & -0.014 \\
$t_{5}$ & 1.238 & $t_{13}$ & -0.072* \\
$t_{6}$ & -0.462 & $t_{14}$ & 0.287* \\
$t_{7}$ & -0.002 & $t_{15}$ & 0.054 \\
$t_{8}$ & 0.175 & \\
& & & \\
$t_{16}$ & 0.052 &$t_{22}$ & -0.037 \\
$t_{17}$ & 0.001 &$t_{23}$ & 0.022 \\
$t_{18}$ & 0.002* & $t_{24}$ & -0.046 \\
$t_{19}$ & -0.041 & $t_{25}$ & 0.010 \\
$t_{20}$ & -0.004* & $t_{26}$ & 0.021 \\
$t_{21}$ & -0.046 & $t_{27}$ & -0.025 \\
& & & \\
$\lambda$ & 1.300  & & \\
$\lambda_1$ & 0.013 & $\lambda_4 $ & 0.010 \\
$\lambda_2$ & -0.025 & $\lambda_5$ & -0.028 \\
$\lambda_3$ & 0.005 & $\lambda_6$ & 0.021 \\
\hline
\hline
\end{tabular*}
\end{table}

Next, we addSOC interactions to the spinless Hamiltonian.
The SOC contribution to the spinor Hamiltonian is expressed as:

\begin{equation}
H_{soc} = \lambda \hat{\bm{L}} \cdot \hat{\bm{S}} 
\label{soc_ab}
\end{equation}
where $\lambda$ is the on-site SOC strength.
This can be expanded further as:
\begin{equation}\label{soc}
H_{soc} = \lambda \frac{\hat{L}_-\hat{S}_++\hat{L}_+\hat{S}_-}{2}+\hat{L}_z\hat{S}_z
\end{equation}
where $L$ is the angular momentum operator on orbitals and $S$ is the angular momentum operator on spin states, $L_{\pm}$ and $S_{\pm}$ are the raising operators defined as $L_x \pm iL_y$ and $S_x \pm iS_y$, respectively.
The on-site SOC strength $\lambda$ has a large effect on inducing band inversion at $\Gamma$ point.
When $\lambda=1.1$ eV, the system experiences a gap-closing phase transition, with exchange of parity character between occupied and unoccupied states.
The VB and CB near the Fermi level will be separated further apart if $\lambda$ is further increased.
We notice a particular feature in DFT bands, which shows a small segment of nonzero curvature in the highest VB near $\Gamma$ just below the Fermi level (Fig. \ref{figbi111band} (b)).
We can obtain this feature with large $\lambda$ at the expense of band mismatches in other parts of the occupied bands.
This issue can be resolved when we introduce a next-nearest neighbor SOC term.
This additional next-nearest neigbor SOC term reflects the strong SOC strength of bismuth and the threefold rotation symmetry of the honeycomb lattice.
The form of this term is expressed as\cite{soc_pg}:

\begin{equation}
H_I=i\sum_{\{m,n\}}\sum_{\{\alpha,\alpha\}}\lambda_{mn}[\sigma_z]_{\alpha\alpha} c^{\dagger}_{m,\alpha} c_{n, \alpha} + h.c. 
\end{equation}
where $H_I$ is the intrinsic SOC originated from the buckling of of the (111) bilayer, $\lambda_{mn}$ is effective SOC strength associated with hopping from orbital $m$ in home cell to orbital $n$ in next-nearest neigbor cell, and $[\hat{\sigma}_z]_{\alpha\alpha}$ is the element $(\alpha,\alpha)$ of the Pauli matrix. As discussed by Kochan et al.\cite{soc_pg}, only diagonal elements in the spin basis are non-zero for intrinsic SOC.  
Combining symmetry considerations and the requirement of Hermiticity, we obtain the intrinsic next-nearest SOC term $\lambda_{mn}$ in the form of a $3\times 3$ matrix with 6 independent parameters (more detailed derivation of this term is given in the Appendix):

\begin{equation}
\lambda_{mn}=
\begin{pmatrix}
\lambda_1 & \lambda_2 &\lambda_3\\
\lambda_2 & \lambda_4 &\lambda_5\\
-\lambda_3 &-\lambda_5 &\lambda_6\\
\end{pmatrix}
\end{equation}

We obtain the parameters from $\lambda_1$ to $\lambda_6$ from first principles again and the on-site SOC term $\lambda$ is tuned to fit the energy bands at $\Gamma$ near the Fermi level.
Therefore the SOC part adds another 7 parameters to our TB model as shown in Table \ref{TB_bilayer_111}.

\begin{table}[h]
\caption{Basis functions of IRs at the $\Gamma$ point for the point group $D_{3d}$, obtained from the projection operator method. \cite{Dresselhaus2008a} }
\label{tab:bilayer_111IR}
\begin{tabularx}{8.5cm}{c|c}
\hline
\hline
IRs & Basis functions \\
\hline
$\Gamma_1^+$ & $\ket{p_z^+}$ \\
$\Gamma_3^+$ & $\{ \ket{p_x^+}, \ket{p_y^+} \}$ \\
$\Gamma_2^-$ & $\ket{p_z^-}$ \\
$\Gamma_3^-$ & $\{\ket{p_x^-}, \ket{p_y^-}\}$ \\
\hline
$\Gamma_4^+ $ & 
$\frac{1}{2}(\ket{p_x^+\uparrow}+i\ket{p_y^+\uparrow} - i\ket{p_x^+\downarrow}-\ket{p_y^+\downarrow})$ \\
$\Gamma_5^+$ &$\frac{1}{2}(\ket{p_x^+\uparrow}+i\ket{p_y^+\uparrow} + i\ket{p_x^+\downarrow}+\ket{p_y^+\downarrow})$ \\
$\Gamma_6^+$ & $\{\ket{\uparrow} , \ket{\downarrow} \}$,\\
&$\{ a_+\ket{p_z^+\uparrow}+\bar{a_+}\frac{1}{\sqrt{2}}(\ket{p_x^+\downarrow}+i\ket{p_y^+\downarrow}),$ \\
& $ a^*_+\ket{p_z^+\downarrow}+\bar{a^*_+}\frac{1}{\sqrt{2}}(\ket{p_x^+\uparrow}-i\ket{p_y^+\uparrow}) \}$ \\
$\Gamma_4^- $ &$\frac{1}{2}(\ket{p_x^-\uparrow}+i\ket{p_y^-\uparrow} - i\ket{p_x^-\downarrow}-\ket{p_y^-\downarrow})$ \\
$\Gamma_5^- $ &$\frac{1}{2}(\ket{p_x^-\uparrow}+i\ket{p_y^-\uparrow} + i\ket{p_x^-\downarrow}+\ket{p_y^-\downarrow})$ \\
$\Gamma_6^-$ & $\{ a_-\ket{p_z^-\uparrow}+\bar{a_-}\frac{1}{\sqrt{2}}(\ket{p_x^-\downarrow}+i\ket{p_y^-\downarrow}),$ \\
& $ a^*_-\ket{p_z^-\downarrow}+\bar{a^*_-}\frac{1}{\sqrt{2}}(\ket{p_x^-\uparrow}-i\ket{p_y^-\uparrow}) \}$ \\
\hline
\hline
\end{tabularx}
\end{table}

This concludes the construction process of our spinless and spinor TB models.
An advantage of these models is that they use minimal parameters to accurately reproduce the electronic structure of the bismuth allotropes studied here.
We now examine the symmetry properties and basis functions of our model to verify this claim.
Fig.~\ref{figbi111band} shows the irreducible representations (IRs) at $\Gamma$ obtained from our TB models and corresponding orbital characters from DFT calculations shown in circles. Table~\ref{tab:bilayer_111IR} summarizes the basis functions of IRs. The notation we use for IRs is consistent with Koster's\cite{koster}. 
The upper part of the table is the basis functions for the spinless case while the bottom part corresponds to the case with spin.
For degenerate states, mixing between basis functions within the degenerate subspace is allowed as long as the two states are related by time reversal (TR) symmetry.

In Fig.~\ref{figbi111band} the highest VB (with IR $\Gamma_3^+$ and basis functions of $\{ \ket{p_x^+}, \ket{p_y^+} \}$) has a double degeneracy protected by the threefold rotation symmetry of the crystal structure and the lowest CB belongs to the $\Gamma_2^-$ IR, which consists of an equal mixing of $p_z$ orbitals from the two atoms in the primitive cell.
These orbital characters are consistent with those from DFT calculations.
Therefore, the eigenstates in our TB models have the correct symmetry properties. 
After taking electron spins into account, the new double group IRs near the Fermi level are: $\Gamma_3^+ \otimes \Gamma_6^+=\Gamma_4^+ \oplus \Gamma_5^+ \oplus \Gamma_6^+$ and $\Gamma_2^- \otimes \Gamma_6^+ = \Gamma_6^-$.
In other words, the product of the basis functions for the highest VB (corresponding to $\Gamma_3^+ $) with spinor basis functions can be decomposed into direct sum of states $\Gamma_4^+$, $\Gamma_5^+$ and $\Gamma_6^+ $, and a product of the lowest CB $\Gamma_2^-$ state with a spinor yields an anti-symmetric $\Gamma_6^-$ state.
As a result of both TR and inversion symmetry, all the eigenstates are doubly degenerate.
In addition, one dimensional IRs $\Gamma_4^\pm $ and $\Gamma_5^\pm$ are related by TR symmetry and form a Kramer pair. 

The band inversion is characterized by the exchange of states between $\Gamma_4^+\oplus\Gamma_5^+$ and $\Gamma_6^-$.
The accompanying exchanging of parity between occupied and unoccupied states gives rise to the topological nontriviality of the system.
The $\mathbb{Z}_2$ topological invariant can be calculated simply by taking parity values at four TR invariant momentum points in the FBZ \cite{Fu2007b}.
Inversion parity eigenvalues of occupied states at Time Reversal Invariant Momenta (TRIM) are listed in Table \ref{z2_bilayer_111}. The $\mathbb{Z}_2$ invariant of 1 calculated from our TB for Bi (111) bilayer is consistent with the values reported in literature using DFT methods\cite{Li2014a}.

\begin{table}[h]
\caption{Inversion eigenvalues of the occupied states at TRIM for the Bi (111) bilayer. The sign in the bracket denotes parity for spinless states. The occupied states are numbered from low energy to high energy and we have used the same indices for degenerate bands.}
\label{z2_bilayer_111}
\begin{tabular*}{\columnwidth}{@{\extracolsep{\fill}} c r c r}
\hline
\hline
TRIM $\backslash$ Occupied States& 1 & 2 & 3 \\
\hline
 (0,0,0) & +(+) & +(+) & $-$(+) \\
 ($\frac{1}{2}$,0,0) & +(+) & $-$($-$) &$-$($-$) \\
(0,$\frac{1}{2}$,0) & +(+) & $-$($-$) &$-$($-$) \\
($\frac{1}{2}$,$\frac{1}{2}$,0) & +(+) & $-$($-$) &$-$($-$) \\
\hline
\hline
\end{tabular*}
\end{table}

\subsection{Bismuthene}
\label{sec:results_bismuthene}

\begin{figure*} [htbp]
    \includegraphics{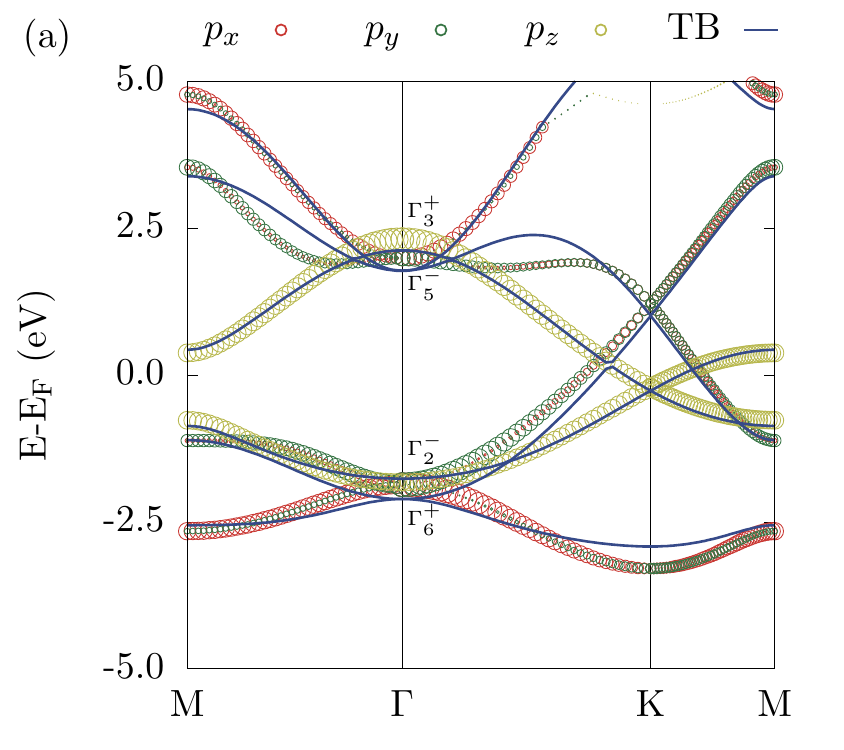}
    \includegraphics{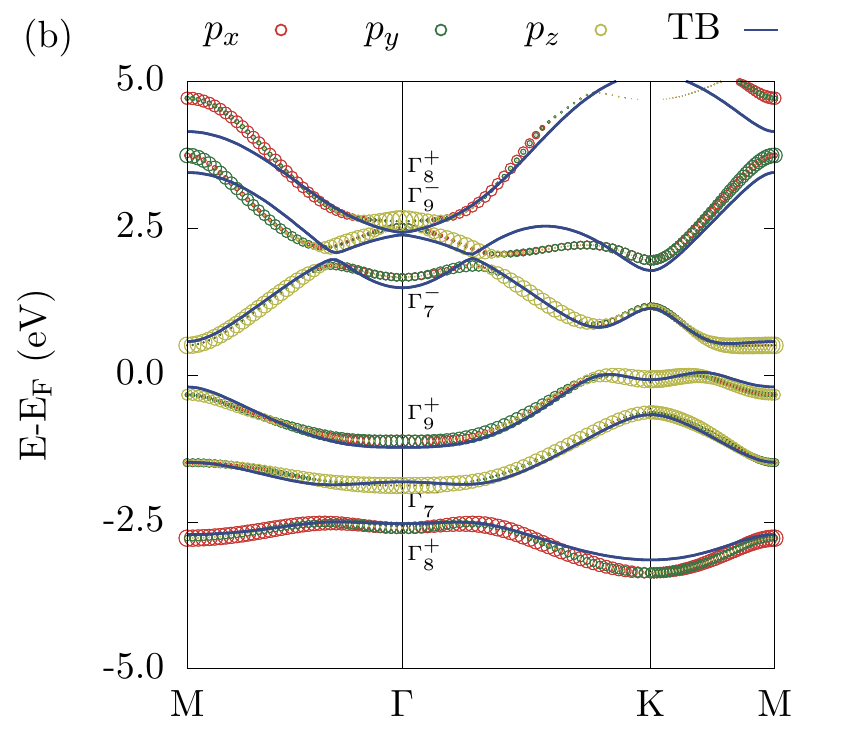}
    \caption{Bismuthene band structure (a) without and (b) with SOC. Energy bands from our TB model are shown as solid dark blue lines. Energy bands from DFT are plotted as circles, with $p_z$ in yellow, $p_x$ in red and $p_y$ in green. The weights of orbitals are represented by the size of circles. }.
    \label{bismuthene_band}
\end{figure*}

Bismuthene is a 2D allotrope of bismuth arranged in a planar honeycomb crystal structure.
The crystal structure has symmetry of space group $P6/mmm$ ($D_{6h}^{1}$, SG191).
The symmetry generators for its point group include identity ($E$), inversion ($P$), twofold rotations along the $z$ direction ($C_{2z}$), and two threefold rotations ($C_{3}$).
By analyzing the orbital character obtained from our DFT calculations (see Fig. \ref{bismuthene_band}) we find that there is negligible hybridization between $s$ and $p$ orbitals, and therefore we choose to include only $p$-orbitals in our TB model of bismuthene.
We have found that the third and fourth NN hoppings have negligible contribution to the band structure and the TB model with next-nearest neighbors already shows satisfactory agreement to DFT results. Therefore, we only include up to next-nearest hoppings in the TB model. First we start with Hamiltonian without spin.
The basis set we take is \{$p_{zBi1}$, $p_{xBi1}$, $p_{yBi1}$, $p_{zBi2}$, $p_{xBi2}$, $p_{yBi2}$\} where the 1 and 2 subscripts distinguish the sublattice atoms.

\begin{table}[h]
\caption{Hopping parameters for bismuthene, up to and including second NNs. The parameter with an asterisk superscript is related with other parameters in the table. }
\label{prm_bismuthene}
\begin{tabular*}{0.5\columnwidth}{@{\extracolsep{\fill}}l l}
\hline
\hline
Parameter & Energy (eV) \\
\hline
$t_{1}$ & -3.919 \\
$t_{2}$ & -3.175 \\
$t_{3}$ & -0.641 \\
$t_{4}$ & 1.943 \\
$t_{5}$ & -0.724 \\
$t_{6}$ & 0.053 \\
$t_{7}$ & -0.152 \\
$t_{8}$ & -0.070 \\
$t_{9}$ & -0.037* \\
$t_{10}$ & -0.091 \\
$\lambda$ & 1.158 \\
\hline
\hline
* $t_9=\sqrt{3}(t_7-t_{10})-t_8$
\end{tabular*}
\end{table}

Similar to the Bi (111) bilayer, the $k$-space Hamiltonian of bismuthene also consists of four sub-blocks:

\begin{equation} \label{Hk-form}
H(\bm{k})=
\begin{pmatrix}
H^{11}(\bm{k}) & H^{12}(\bm{k}) \\
H^{21}(\bm{k}) & H^{22}(\bm{k})\\
\end{pmatrix}
\end{equation}
The form of the sub-blocks are:

\begin{equation}
H^{11}(\bm{k}) =
\begin{pmatrix}
H_{11}& 0 & 0 \\
0 & H_{22} & H_{23} \\
0 & H_{32} & H_{33} \\
\end{pmatrix}
\end{equation}

\begin{equation}
H^{12}(\bm{k}) =
\begin{pmatrix}
H_{14} & 0 & 0 \\
0 & H_{25} &H_{26} \\
0 & H_{35} &H_{36} \\
\end{pmatrix}
\end{equation}

Again, we only need to solve for $H^{11}(\bm{k})$ and $H^{12}(\bm{k})$ since the other two sub-blocks can be obtained by inversion and Hermitan adjoint.
The $H^{ij}(\bm{k})$ of bismuthene has the same components $H_{ij}(\bm{k})$ as the Bi (111) bilayer except those zero elements in $H^{11}(\bm{k})$ and $H^{22}(\bm{k})$, because bismuthene has all the symmetry element generators of Bi (111) bilayer with an extra horizontal mirror symmetry $\sigma_h$ element.
The general form of the on-site hopping term $H_{ii}(\bm{k})$ is:

\begin{equation}\label{diag_1}
H_{ii}(\bm{k})=\epsilon_i+2h_{ii}^{(1)}[\cos(\bm{k}\cdot\bm{r_4})+\cos(\bm{k}\cdot\bm{r_6})]+2h_{ii}^{(2)}\cos(\bm{k}\cdot\bm{r_5})
\end{equation}
where $\epsilon_1$, $\epsilon_2$, $\epsilon_3$=$t_1, t_2, t_2$; $h_{11}^{(1)}$, $h_{22}^{(1)}$, $h_{33}^{(1)}$ = $t_6, t_{7}, t_{10}$; $h_{11}^{(2)}$, $h_{22}^{(2)}$, $h_{33}^{(2)}$=$t_6, (\frac{3}{2}t_{10}-\frac{1}{2}t_{7}), (\frac{3}{2}t_{7}-\frac{1}{2}t_{10})$ respectively.
The only off-diagonal term $H_{23}(\bm{k})$ in $H^{11}(\bm{k})$ has the form:

\begin{equation}\label{offdiag_23}
\begin{aligned}
H_{23}(\bm{k})=2ih_{23}^{(1)}\sin(\bm{k}\cdot\bm{r_5})&+h_{23}^{(2)}(e^{i\bm{k}\cdot\bm{r_4}}-e^{-i\bm{k}\cdot\bm{r_6}})\\
&+h_{23}^{(3)}(e^{-i\bm{k}\cdot\bm{r_4}}-e^{i\bm{k}\cdot\bm{r_6}})
\end{aligned}
\end{equation} 
where the constants $h_{23}^{(1)}=(\frac{1}{2}t_{9}-\frac{1}{2}t_{8})$, $h_{23}^{(2)} = t_{8}$, and $h_{23}^{(3)} = t_{9}$. For off-diagonal hopping sub-block $H^{12}(\bm{k})$, the components are given by:

\begin{equation}\label{offdiag_14}
H_{14}(\bm{k}) = h_{14}^{(1)}( e^{i\bm{k}\cdot\bm{r_1}}+e^{i\bm{k}\cdot\bm{r_2}}+e^{i\bm{k}\cdot\bm{r_3}})
\end{equation}

\begin{equation}\label{offdiag_25}
H_{25}(\bm{k}) = h_{25}^{(1)}e^{i\bm{k}\cdot\bm{r_1}}+h_{25}^{(2)}e^{i\bm{k}\cdot\bm{r_2}}+h_{25}^{(2)}e^{i\bm{k}\cdot\bm{r_3}}
\end{equation}

\begin{equation}\label{offdiag_26}
H_{26}(\bm{k}) = h_{26}^{(1)}e^{i\bm{k}\cdot\bm{r_2}}-h_{26}^{(1)}e^{i\bm{k}\cdot\bm{r_3}}
\end{equation}

\begin{equation}\label{offdiag_35}
H_{35}(\bm{k}) = h_{35}^{(1)}e^{i\bm{k}\cdot\bm{r_2}}-h_{35}^{(1)}e^{i\bm{k}\cdot\bm{r_3}}
\end{equation}

\begin{equation}\label{offdiag_36}
H_{36}(\bm{k}) = h_{36}^{(1)}e^{i\bm{k}\cdot\bm{r_1}}+h_{36}^{(2)}e^{i\bm{k}\cdot\bm{r_2}}+h_{36}^{(2)}e^{i\bm{k}\cdot\bm{r_3}}
\end{equation}
Again, $h_{14}^{(1)}=t_3$, $h_{25}^{(1)}=t_4$, $h_{25}^{(2)}=(\frac{1}{4}t_4 + \frac{3}{4} t_5)$,$h_{35}^{(1)}$ = $h_{26}^{(1)}=(-\frac{\sqrt{3}}{4}t_4+\frac{\sqrt{3}}{4}t_5)$, $h_{36}^{(1)}=t_5$ and $h_{36}^{(2)}=(\frac{3}{4}t_4 + \frac{1}{4}t_5)$. Finally, we obtain 9 independent hopping parameters as shown in Table \ref{prm_bismuthene}.

We then include the effect of SOC in the TB Hamiltonian by duplicating the basis set to introduce a spin degree of freedom and adding an on-site SOC term $H_{soc}$.
The SOC strength for bismuthene ($\lambda$) is taken directly from our $\it{ab}$ ${initio}$ calculation as defined in Eq. \ref{soc_ab}.
This approach is different from that we used for a Bi (111) bilayer where we have added the next-nearest SOC terms to capture the non-negligible buckling effect on the SOC.
In comparison, we found that for bismuthene the on-site atomic SOC term alone can well reproduce the band structure.
The $\lambda$ value for bismuthene is presented in Table \ref{prm_bismuthene}.

We check the symmetry properties of bismuthene in our TB model by identifying the IRs of the eigenstates at the $\Gamma$ point and projecting them onto their corresponding basis functions, listed in Table \ref{tab:bismuthene}.
We compare these IRs with projected orbital characters predicted by DFT as shown in Fig. \ref{bismuthene_band}.
Without SOC, the eigenstates formed by spinless $p$-orbitals can be categorized into representations $\Gamma_2^-$, $\Gamma_3^+$, $\Gamma_6^+$ and $\Gamma_5^-$. $\Gamma_3^+$ and $\Gamma_6^+$ states correspond to symmetric states formed by out-of-plane $p$-orbitals and in-plane $p$-orbitals respectively. $\Gamma_2^-$ and $\Gamma_5^-$ correspond to anti-symmetric states. In the Fig. \ref{bismuthene_band}, it can be seen that the orbital characters from TB model are consistent with those from DFT calculations. 
 
To analyze the symmetry of the Bloch states with SOC, we introduce spinor representation $\Gamma_7^+$ and generate IRs of spinful eigenstates from previous IRs of spinless states by again taking the direct product with spinor and decomposing onto IRs of the double group for occupied states: $\Gamma_6^+ \otimes \Gamma_7^+ = \Gamma_9^+ \oplus \Gamma_8^+$, $\Gamma_2^- \otimes \Gamma_7^+ = \Gamma_7^-$.
The IRs for states in the CB are obtained in a similar way as $\Gamma_7^-$, $\Gamma_8^+$ and $\Gamma_9^-$.
To label the eigenstates, we calculate symmetry eigenvalues at $\Gamma$ point and compare with the $D_{6h}$ double group character table listed in Table XII and XIII in the Appendix.
 
SOC has several effects on the electronic states of bismuthene.
First we notice the degeneracy corresponding to the $\Gamma_6^+$ state and $\Gamma_5^-$ state is lifted.
Band exchange happens between the upper two bands in the VB and another two high-energy bands in the CB.
Even though SOC exchanges the parity of these two VB states, it does not affect the topological property of bismuthene because there is no exchange between unoccupied and occupied states.
We also see that with SOC the electronic structure of bismuthene undergoes a transition from semimetallic phase to an insulator phase after a band gap opens near the $K$ point.

We confirm this further by calculating the $\mathbb{Z}_2$ invariant using parity values at TRIM listed in Table \ref{z2_bismuthene}. We have numbered bands from low energy to high energy and neglected spin degeneracy for brevity.
The result that bismuthene has $\mathbb{Z}_2=0$, is consistent with the 
previous studies.\cite{Huang2013a}

\begin{table}[h]
\caption{Basis functions of IRs at $\Gamma$ point (point group $D_{6h})$, obtained from projection operator method. \cite{Dresselhaus2008a}. There are two sets of $p$-orbital combinations allowed for $\Gamma^{8+}$ and $\Gamma^{7-}$ IRs. The bands with these two IRs consist of linear combination of the in-plane and out-of-plane $p$-orbitals, in a similar way to $\Gamma^{6+}$ and $\Gamma^{6-}$ IRs from $D_{3d}$ group as listed in Table~\ref{tab:bilayer_111IR}. The two basis functions within the two dimensional sub-space of these IRs form a Kramer pair conjugated to each other.}
\label{tab:bismuthene}
\begin{tabular*}{\columnwidth}{@{\extracolsep{\fill}}l l}
\hline
\hline
Irreducible Rep. & Basis functions \\
\hline
$\Gamma_3^+$ & $\ket{p_z^+}$ \\
$\Gamma_6^+$ & $\{\ket{p_x^+} , \ket{p_y^+}\}$ \\
$\Gamma_2^-$ & $\ket{p_z^-}$ \\
$\Gamma_5^-$ & $\{\ket{p_x^-}, \ket{p_y^-}\}$ \\
\hline
$\Gamma_7^+$ & $\{\ket{\uparrow} , \ket{\downarrow} \}$ \\
$\Gamma_8^+$ & $\{\frac{1}{\sqrt{2}} (\ket{p_x^+ \uparrow} - i\ket{p_y^+\uparrow}) , \frac{1}{\sqrt{2}} (\ket{p_x^+\downarrow} + i\ket{p_y^+\downarrow}) \}$ \\
&$\{\ket{p_z^+ \downarrow} , \ket{p_z^+ \uparrow} \}$ \\
$\Gamma_9^+$ & $\{\frac{1}{\sqrt{2}} (\ket{p_x^+ \downarrow} - i\ket{p_y^+ \downarrow}) , \frac{1}{\sqrt{2}} (\ket{p_x^+ \uparrow} + i\ket{p_y^+ \uparrow}) \}$ \\
$\Gamma_7^-$ & $\{\frac{1}{\sqrt{2}} (\ket{p_x^- \uparrow} - i\ket{p_y^-\uparrow}) , \frac{1}{\sqrt{2}} (\ket{p_x^-\downarrow} + i\ket{p_y^-\downarrow}) \}$ \\
&$\{\ket{p_z^- \downarrow} , \ket{p_z^- \uparrow} \}$ \\
$\Gamma_9^-$ & $\{\frac{1}{\sqrt{2}} (\ket{p_x^- \downarrow} - i\ket{p_y^- \downarrow}) , \frac{1}{\sqrt{2}} (\ket{p_x^- \uparrow} + i\ket{p_y^- \uparrow}) \}$ \\
\hline
\hline
\end{tabular*}
\end{table}

\begin{table}[h]
\caption{Inversion eigenvalues of occupied states at TRIM for bismuthene. The sign in the bracket denotes parity for spinless states}
\label{z2_bismuthene}
\begin{tabular*}{\columnwidth}{@{\extracolsep{\fill}} c r c r}
\hline
\hline
TRIM $\backslash$ Occupied States& 1 & 2 & 3 \\
\hline
(0,0,0) & +(+) & $-$(+) & $-$($-$) \\
($\frac{1}{2}$,0,0) & $-$($-$) & +(+) &+(+) \\
(0,$\frac{1}{2}$,0) & $-$($-$) & +(+) &+(+) \\
($\frac{1}{2}$,$\frac{1}{2}$,0) & $-$($-$) & +(+) &+(+) \\
\hline
\hline
\end{tabular*}
\end{table}

\subsection{Bi (110) bilayer}
\label{sec:results_110_bilayer}

\begin{figure*}[htbp]
  \includegraphics{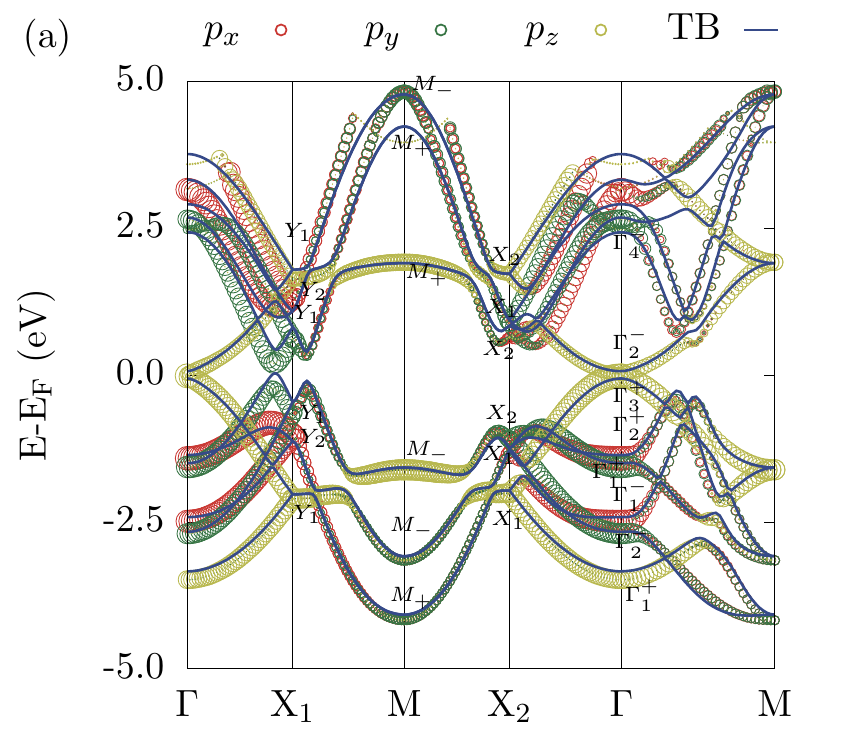}
  \includegraphics{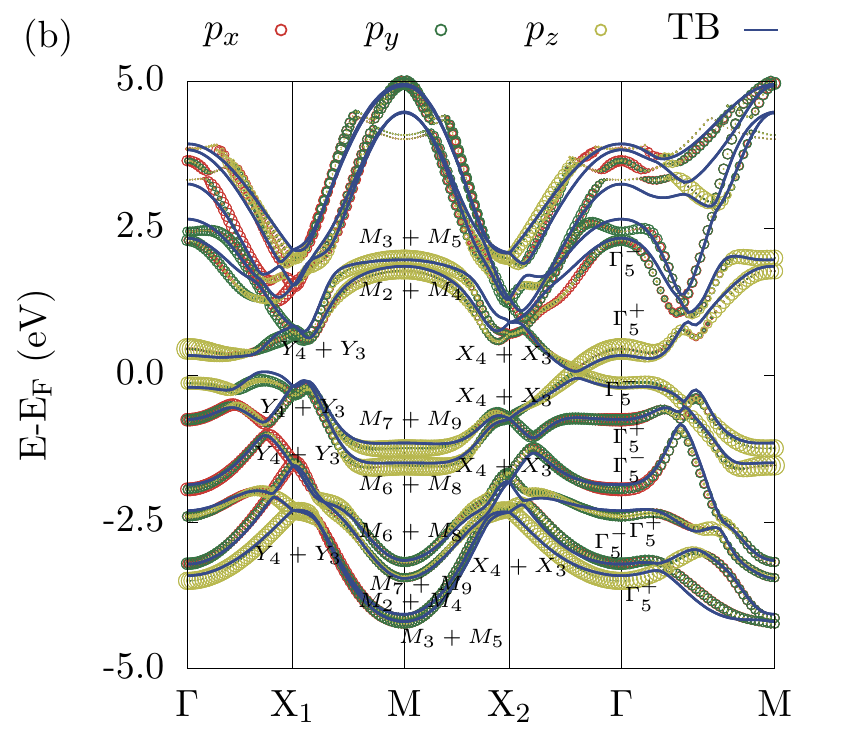}
  \caption{Bi (110) bilayer band structure (a) without and (b) with SOC. Energy bands from our TB model are shown as solid dark blue lines. Energy bands from DFT are plotted as circles, with $p_z$ in yellow, $p_x$ in red and $p_y$ in green. The weights of orbitals are represented by the size of circles. }
  \label{Bi110}
\end{figure*}

A Bi (110) bilayer consists of two layers of planar bismuth. Upon relaxation, we find this 2D allotrope has a non-symmorphic crystal symmetry unlike the (110) surface of bulk bismuth, which has been reported experimentally \cite{exp110}. The absence of buckling in each planar layer leads to a crystal symmetry of space group $Pmna$ ($D_{2h}^{7}$, SG53). 
The symmetry generators for this group include two twofold rotations $C_{2y}$, $C_{2z}$ and two corresponding mirror operations $\sigma_{2y}$, $\sigma_{2z}$ with a glide translation vector of ($\frac{1}{2}, \frac{1}{2}, 0$).
The presence of glide plane symmetry usually results in additional band degeneracies in the electronic structure.
The projected orbital characters from our DFT calculations show that electronic bands of Bi (110) bilayer are mainly from $p$-orbitals (see Fig. \ref{Bi110}). The contribution of $s$-orbitals is to energy bands far below the Fermi level and in this case isolated from the rest of the band structure. Therefore we only consider $p$-orbitals as basis functions in our TB model for Bi (110) bilayer.

It is more convenient to use neighboring cells instead of neighboring atoms in the construction of TB models for Bi (110) bilayer since it has a low-symmetry crystal structure and an accurate description of its electronic structure requires inclusion of more neighboring atoms compared to the other two allotropes. 
In this way, we have used interactions between orbitals in the home cell and orbitals in all eight neighbor cells around it as shown in Fig. \ref{atom_schematics} (f).
 
As before, we start by constructing the spinless Hamiltonian. 
The real-space Hamiltonian matrix describes hopping between the home cell and neighbor cells at $\bm{R}$. The entries of the Hamiltonian can be divided into 16 sub-blocks to represent hopping between four inequivalent atoms in the primitive cells. 
Therefore, we analyze the form of the real-space Hamiltonian in the unit of a sub-block. For hopping in the home cell $H_{[000]}$, there are 4 independent sub-blocks in the Hamiltonian: $H^{11}_{[000]}$, $H^{12}_{[000]}$, $H^{13}_{[000]}$ and $H^{14}_{[000]}$. The other sub-blocks can be derived from these independent sub-blocks by applying $\sigma_{(001)}$, inversion, and $C_2^{[001]}$ operations. Symmetries also restrict the number of independent parameters in each sub-block. Finally we can write the hopping in the home cell as:


\begin{center}
$
H_{[000]}=\begin{pmatrix}
H^{11}_{[000]} & H^{12}_{[000]} & H^{13}_{[000]} & H^{14}_{[000]} \\
& \sigma_{z}(H^{11}_{[000]}) & C_{2y}(H^{14}_{[000]}) & 0 \\
& & i( H^{11}_{[000]}) & C_{2x}(H^{12}_{[000]}) \\
& h.c. && C_{2x}(H^{11}_{[000]})
\end{pmatrix}
$
\end{center}
Next, we consider hoppings from the home cell to cell $\bm{R}=[100]$. There are only two independent sub-blocks $H^{11}_{[100]}$ and $H^{24}_{[100]}$:

\begin{center}
$
H_{[100]}=
\begin{pmatrix}
H^{11}_{[100]} & 0 & 0 & 0\\
\sigma_{x}(H^{12\dagger}_{[000]}) & \sigma_{z}(H^{11}_{[100]}) & \sigma_{x}(H^{23}_{[000]}) & H^{24}_{[100]} \\
0&0 & C_{2x}(H^{11}_{[100]}) & 0 \\
\sigma_{y}(H^{14}_{[000]})& C_{2x}(H^{24}_{[100]})& i(H^{21}_{[000]})& \sigma_y(H^{11\dagger}_{[100]})\\
\end{pmatrix}
$
\end{center}

Similarly for hopping to cell $\bm{R}=[010]$, $\bm{R}=[110]$ and $\bm{R}=[1\bar{1}0]$, we have:

\begin{center}
$
H_{[010]}=
\begin{pmatrix}
H^{11}_{[010]} & 0 & 0 & H^{14}_{[010]} \\
C_{2y}(H^{12}_{[000]})& C_{2y}(H^{11}_{[010]})& C_{2y}(H^{14}_{010]})& C_{2y}(H^{13}_{[000]})& \\
0&0 & i(H^{11}_{[010]})^\dagger & C_{2z}(H^{21}_{[000]}) \\
0& 0&0& \sigma_z(H^{33}_{[010]})
\end{pmatrix}
$
\end{center}

\begin{center}
$
H_{[110]}=
\begin{pmatrix}
H^{11}_{[110]} & 0 & 0 & 0 \\
\sigma_z(H^{12}_{[000]})&\sigma_z(H^{11}_{[110]})& \sigma_x(H^{23}_{[010]})& 0& \\
\sigma_y(H^{24}_{[100]})&0 & i(H^{11}_{[110]})^\dagger & 0 \\
0& 0&0& \sigma_z(H^{33}_{[110]})
\end{pmatrix}
$
\end{center}

\begin{center}
$
H_{[1\bar{1}0]}=
\begin{pmatrix}
C_{2x}(H^{33}_{[110]}) & 0 & C_{2x}(H^{31}_{[110]}) & 0 \\
0 & C_{2x}(H^{44}_{[110]}) & 0 & 0 \\
0 & 0 & C_{2x}(H^{11}_{[110]}) & 0 \\
C_{2x}(H^{23}_{[110]}) & 0 & C_{2x}(H^{21}_{[110]}) & C_{2x}(H^{22}_{[110]})
\end{pmatrix}
$
\end{center}
The rest of the hoppings to cells at $\bm{R}=[\bar{1}00]$, $\bm{R}=[0\bar{1}0]$, $\bm{R}=[\bar{1}\bar{1}0]$ and $\bm{R}=[\bar{1}10]$ can be easily obtain by the Hermitian condition $H(-\bm{R})=H(\bm{R})^\dagger$. SOC is added by duplicating the basis and directly evaluating matrix elements $\bra{w_i}\lambda \hat{L}\cdot\hat{S}\ket{w_j}$ where $i$ and $j$ denote the Wannier orbitals in the home cell and $\lambda$ is the on-site SOC strength. Table \ref{tab:bilayer_110} summarizes all the hopping parameters for the Bi(110) bilayer.
 
 \begin{table}[htbp]
\caption{Hopping parameters for the Bi (110) bilayer, up to and including the nearest cells to the home cell.}
\label{tab:bilayer_110}
\begin{tabular*}{\columnwidth}{@{\extracolsep{\fill}} c r c r}
\hline
\hline
Parameter & Energy (eV) & Parameter & Energy (eV) \\
\hline
$t_{1}$ & -2.375 & $t_{19}$ & 0.063 \\
$t_{2}$ & -2.075 & $t_{20}$ & 0.051 \\
$t_{3}$ & -2.183 & $t_{21}$ & -0.111\\
$t_{4}$ & -0.080 &$t_{22}$ & 0.038 \\
$t_{5}$ & 1.766 & $t_{23}$ & 0.041 \\
$t_{6}$ & -0.501 & $t_{24}$ & -0.084 \\
$t_{7}$ & -0.453 & $t_{25}$ & -0.301\\
$t_{8}$ & -0.284 & $t_{26}$ & 0.376 \\
$t_{9}$ & -0.573 &$t_{27}$ & 0.657 \\
$t_{10}$ & 0.847 &$t_{28}$ & 0.019\\
$t_{11}$ & 0.529&$t_{29}$ & 0.016\\ 
$t_{12}$ & 1.232 &$t_{30}$ & 0.774\\
$t_{13}$ & -0.031 &$t_{31}$ & 0.036\\
$t_{14}$ & 0.026 &$t_{32}$ & 0.072\\
$t_{15}$ & 0.038 &$t_{33}$ & 0.048\\
$t_{16}$ & 0.064 &$t_{34}$ & 0.016\\
$t_{17}$ & 0.031 &$t_{35}$ & 0.014\\
$t_{18}$ & -0.056 &$\lambda$ & 1.240\\
\hline
\hline
\end{tabular*}
\end{table}
 
We now examine the symmetry properties of electronic bands near the Fermi level.
For spinless eigenstates, the electronic bands at $\Gamma$ consist of mixing between the sets of $p_y$ and $p_z$ orbitals, while mixing of $p_x$ orbitals is disallowed, as shown in Table \ref{Gnsoc}. After SOC is included, the orbital characters at $\Gamma$ are a mixture of all $p$-orbitals. The effect of SOC also induces a band inversion at $\Gamma$ between the two bands at either side of the Fermi level, implying a gap-closing topological phase transition. 

\begin{table}[h]
\caption{Basis functions of IRs at $\Gamma$ point for two states near Fermi level, obtained from projection operator method. \cite{Dresselhaus2008a} }
\label{Gnsoc}
\begin{tabular*}{\columnwidth}{@{\extracolsep{\fill}}l l}
\hline
\hline
Irreducible Rep. & Basis functions \\
\hline
$\Gamma_2^-$ & $\frac{1}{2}(\ket{p_{y1}} - \ket{p_{y2}} +\ket{p_{y3}}-\ket{p_{y4}})$ \\
& $\frac{1}{2}(\ket{p_{z1}} + \ket{p_{z2}} +\ket{p_{z3}}+\ket{p_{z4}})$ \\
$\Gamma_3^+$ & $\frac{1}{2}(\ket{p_{y1}} - \ket{p_{y2}}-\ket{p_{y3}}+\ket{p_{y4}})$ \\
& $\frac{1}{2}(\ket{p_{z1}} + \ket{p_{z2}}-\ket{p_{z3}}-\ket{p_{z4}})$ \\
\hline
\hline
\end{tabular*}
\end{table}

\begin{table}[h]
\caption{Basis functions of IRs at $Y$ point, obtained from projection operator method. \cite{Dresselhaus2008a} }
\label{tab:Ynsoc}
\begin{tabular*}{\columnwidth}{@{\extracolsep{\fill}}l l}
\hline
\hline
Irreducible Rep. & Basis functions \\
\hline
$Y_1$ & \{$\frac{1}{\sqrt{2}}(\ket{p_{y1}} - \ket{p_{y3}})$ ,$\frac{1}{\sqrt{2}}(\ket{p_{y2}} - \ket{p_{y4}})$ \} \\
& \{$\frac{1}{\sqrt{2}}(\ket{p_{z1}} - \ket{p_{z3}})$ ,$\frac{1}{\sqrt{2}}(\ket{p_{z2}} - \ket{p_{z4}})$ \} \\

$Y_2$ & \{$\frac{1}{\sqrt{2}}(\ket{p_{x1}} - \ket{p_{x3}})$ ,$\frac{1}{\sqrt{2}}(\ket{p_{x2}} - \ket{p_{x4}})$ \} \\
\hline
\hline
\end{tabular*}
\end{table}

The most notable feature in the electronic bands of the Bi (110) bilayer is the additional degenerate states at $X$ and $Y$ due to presence of glide mirror symmetries. For a spinless Hamiltonian, the bands at $X$ and $Y$ are doubly degenerate. The degenerate state at $X$ and $Y$ are different in terms of orbital characters. Moreover, at $X$, the bands consist of all $p$-orbitals, while at Y the mixing between $p_x$-orbitals and other $p$-orbitals is disallowed. The bands at $X$ are doubly degenerate with $\pm C_{2x}$ eigenvalues, while the doubly degenerate bands at $Y$ are distinguished by opposite $\sigma_{x}$ eigenvalues. The band representations at $X$ and $Y$ have an additional dimension compared to that at $\Gamma$ as a consequence of the effect of non-symmorphic glide mirror symmetries. This additional dimension in band representations is an important requirement for the existence of symmetry-protected band degeneracies. 
 
Taking spins into consideration, the orbital characters at $X$ and $Y$ are from mixing of the $p$-orbitals across all atoms. The 2-fold band degeneracy imposed by the glide mirror symmetries is still retained. Then combining with TR and inversion symmetry, the bands will become fourfold degenerate with SOC included. The representations of these fourfold degenerate bands are related to each other by TR and $\sigma_{z}$ symmetry. Finally, we calculate inversion eigenvalues of the occupied states and list them in Table \ref{z2_110}. $\mathbb{Z}_2$=1 indicates Bi (110) bilayer is a topological insulator with non-symmorphic symmetry \cite{Lu2015a}.

\begin{table}[h]
\caption{Inversion eigenvalues of occupied states at TRIM for the Bi (110) bilayer. The sign in the bracket denotes parity for non-SOC states.}
\label{z2_110}
\begin{tabular*}{\columnwidth}{@{\extracolsep{\fill}} c r c r c r c}
\hline
\hline
TRIM $\backslash$ Occupied States& 1 & 2 & 3 &4 &5 &6\\
\hline
(0,0,0) & +(+) & $-$($-$) & +($-$) & $-$(+) & +(+) & $-$(+) \\
($\frac{1}{2}$,0,0) & +(+) & $-$($-$) &+(+) & $-$($-$) &+(+) & $-$($-$)\\
(0,$\frac{1}{2}$,0) & +(+) & $-$($-$) &+(+) & $-$($-$) &+(+) & $-$($-$)\\
($\frac{1}{2}$,$\frac{1}{2}$,0) & +(+) & +(+) &$-$($-$)&$-$($-$)&$-$($-$)&$-$($-$)\\
\hline
\hline
\end{tabular*}
\end{table}

\section{Conclusion}
\label{sec:conclusion}

In conclusion, we have constructed tight binding models for three different allotropes of 2D bismuth, namely, the Bi (111) bilayer, bismuthene and the Bi (110) bilayer, by projecting Bloch states calculated from DFT onto localized atomic orbitals as Wannier functions. 
We have utilized a minimal set of independent hopping parameters in the TB model to represent the energy bands near the Fermi level. These TB models can accurately reproduce the band topology of these three allotropes compared to previous semi-empirical TB models.

In all cases considered here, we find that the crystalline symmetry plays an important role in constructing Wannier-based TB models.
We have tailored the form of the TB model to meet the symmetry constraints imposed on each type of bismuth allotrope.
These symmetry constraints can eliminate the numerical errors in the Wannier Hamiltonian, leading to a significant reduction in the number of independent TB parameters.
Symmetrized TB models can faithfully recover the band representations near the Fermi level for each bismuth allotrope and also help us to understand the physical origin of specific features in the band structures.

In summary, we have shown that TB models of two-dimensional bismuth allotropes can be effectively derived from the Wannier basis. 
This symmetry-based approach could be conveniently applied to other topologically nontrivial two-dimensional materials such as antimonene and arsenene \cite{doi:10.1021/acs.nanolett.9b02444,C5NR05006E}. 
Another advantage of these simplified but accurate models is that they can be easily modified to incorporate external factors such as strain, electric and magnetic fields \cite{doi:10.1021/acs.jpclett.7b00222,PhysRevLett.122.036401}. 
Moreover, these models can be a cornerstone for designing new topological structures exhibiting exotic quantum transport properties. 
For example, the tight-binding models proposed by Su, Schrieffer and Heeger \cite{PhysRevLett.42.1698} have inspired experimental realization of stable topological quantum states in graphene nanoribbons \cite{groning2018engineering}. 
These models could also be effective in investigating structure complexities in topological systems, e.g. interactions with defects and disorders \cite{zangeneh2020disorder}. 
Therefore we expect our simplified and symmetrized tight-binding models could play a critical role in multi-scale modeling of low-dimensional topological structures, potentially offering an accurate picture of device physics at lowered computational cost \cite{KLYMENKO2021107676}.

\section{Acknowledgements}

The authors acknowledge the support of the Australian Research Council Centre of Excellence in Future Low-Energy Electronics Technologies. (CE170100039).
The authors also acknowledge the support from Australian National Computing Infrastructure and Pawsey Supercomputing Centre.
MVK acknowledges the support of the Australian Research Council Centre of Excellence in Exciton Science (CE170100026).

\section{Appendix}

\appendix

\subsection{Symmetry constraints on a real space Hamiltonian}

\subsubsection{Spinless real space Hamiltonian of a Bi (111) bilayer}

Here we present how to generate a real space Hamiltonian from the interaction effect of symmetries operations on atomic orbitals and lattice sites. The crystal structure of a Bi (111) bilayer has the $D_{3d}$ point group symmetry. The symmetry constraints on the real space basis function include Identity (E), inversion (I), threefold rotation ($C_3$) and mirror symmetry ($\sigma$). We can write down the matrix representation of these symmetry operations on a set of p orbital basis $\{p_z, p_x, p_y\}$ from the same atom:

\begin{align*}
    D(E) &=
    \begin{pmatrix}
    1& 0&0\\
    0& 1&0\\
    0&0&1
    \end{pmatrix}\\    
    D(i) &=
    \begin{pmatrix}
    -1& 0&0\\
    0& -1&0\\
    0&0&-1
    \end{pmatrix}\\
    D(C_3^+) &=
    \begin{pmatrix}
    1& 0&0\\
    0& -\frac{1}{2}&\frac{\sqrt{3}}{2}\\
    0&-\frac{\sqrt{3}}{2}&-\frac{1}{2}
    \end{pmatrix}\\
    D(\sigma_{110}) &=
    \begin{pmatrix}
    1& 0&0\\
    0& 1&0\\
    0&0&-1
    \end{pmatrix}
\end{align*}

Other symmetry elements can be generated using:
\begin{equation*}
S_6^+ = iC_3^- 
\end{equation*}
\begin{equation*}
S_6^- = iC_3^+ 
\end{equation*}
\begin{equation*}
\sigma_{010} = \sigma_{110}C_3^-
\end{equation*}
\begin{equation*}
\sigma_{100} = \sigma_{110}C_3^+
\end{equation*}
\begin{equation*}
C_2^{010} = i\sigma_{010}
\end{equation*}
\begin{equation*}
C_2^{100} = i\sigma_{100}
\end{equation*}
\begin{equation*}
C_2^{110} = i\sigma_{110}
\end{equation*}

We also introduce matrices to describe the symmetry operations on lattice sites. In the primitive cell of Bi (111) bilayer, there are two atoms. We can therefore write these operations as 2$\times$2 matrices in the basis of lattice sites {1,2} as:

\begin{align*}
    D(E)=
    \begin{pmatrix}
    1&0\\
    0&1\\
    \end{pmatrix}\\
    D(C_3^+)=
    \begin{pmatrix}
    1&0\\
    0&1\\
    \end{pmatrix}\\
    D(i)=
    \begin{pmatrix}
    0&1\\
    1&0\\
    \end{pmatrix}\\
    D(\sigma_{110})=
    \begin{pmatrix}
    0&1\\
    1&0\\
    \end{pmatrix}
\end{align*}

We can readily obtain the form of real space spinless Hamiltonian by Equation 5. The $D(g)$ matrices in the Equation 5 are 6$\times$6 matrices as a result of the tensor product between orbital representation matrices and lattice site representation matrices. In addition, the real space Hamiltonian should be Hermitian:

\begin{equation} 
H_{\bm{R}} = H_{\bm{-R}}^\dagger,
\end {equation}

where $\bm{R}$ is a lattice vector. This can be derived from a Fourier transform of the Bloch Hamiltonian. 

In the following, we give the form of different real space Hamiltonians of a Bi (111) bilayer by considering up to $4^{th}$ atomic neighbors in detail. Each real space Hamiltonian consists of 4 3$\times$3 sub-blocks corresponding to four possible ways of hoppings between two lattice sites. Therefore we can analyze different atomic neighbor hoppings by focusing on specific sub-blocks of a real space Hamiltonian. The hopping terms in all real space Hamiltonians are significantly constrained by symmetry operations. These terms are given in Table \ref{TB_bilayer_111}.

First, for investigating the on-site and the nearest atomic NNs, we derive the real space Hamiltonian for the home cell [000]:

\begin{align}
    H_{[000]} =
    \begin{pmatrix}
    t_1&0&0&t_3&t_4&0\\
    0&t_2&0&t_4&t_5&0\\
    0&0&t_2&0&0&t_6\\
    t_3&t_4&0&t_1&0&0\\
    t_4&t_5&0&0&t_2&0\\
    0&0&t_6&0&0&t_2
    \end{pmatrix}
\end{align}

The diagonal sub-blocks are on-site hoppings while the off-diagonal sub-blocks are hopping between lattice sites in neighboring cells and the home cell. In the diagonal sub-blocks, $C_3$, $\sigma$, and Hermiticity lead to non-zero hopping terms in the diagonal entries of $H_{[000]}$. As the two lattice sites are related by inversion symmetry, the two diagonal sub-blocks should have the same form, resulting in the only independent parameters in the diagonal sub-blocks to being $t_{1}$ and $t_{2}$. Similarly, for the off-diagonal sub-blocks, the symmetry constraints also limits the number of independent parameters to four. Consequently, $H_{[000]}$ can be written with only six independent hopping terms.


We now analyze the form of next-nearest atomic neighbor hopping terms by examining the interactions between the home cell $\bm{R}=[000]$ and the cell at $\bm{R}=[100]$. There are also four types of hopping terms here: hopping from lattice site 1 (2) in home cell to lattice site 1 (2) in the cell at $\bm{R}=[100]$. The real space Hamiltonian of cell $\bm{R}=[100]$ is expressed in the form:

\begin{align}
    H_{[100]} =
    \begin{pmatrix}
    t_7&t_8&t_9&0&0&0\\
    t_{10}&t_{11}&t_{12}&0&0&0\\
    t_{13}&t_{14}&t_{15}&0&0&0\\
    0&0&0&t_7&t_{10}&t_{13}\\
    0&0&0&t_8&t_{11}&t_{14}\\
    0&0&0&t_9&t_{12}&t_{15}\\
    \end{pmatrix}
\end{align}

By applying mirror symmetry $\sigma_{(100)}$ and Hermiticity, further constraints can be found on $H_{[100]}$:

\begin{equation}
t_{10} = -\frac{1}{2}t_8 - \frac{\sqrt{3}}{2}t_9 
\end{equation}
\begin{equation}
t_{13} = \frac{\sqrt{3}}{2}t_8 + \frac{1}{2}t_9 
\end{equation}
\begin{equation}
t_{14} = \sqrt{3}t_{11} - t_{12} - \sqrt{3}t_{15} 
\end{equation}

Hence there are only six independent parameters in the next-nearest neighbor part of $H_{[100]}$. The off-diagonal blocks are zero since we only include the terms for next-nearest atomic neighbors here.

As for $3^{rd}$ atomic neighbor interactions, we use the off-diagonal part of the real space Hamiltonian of cell $\bm{R}=[110]$, which has the form:

\begin{align}
    H_{[110]} =
    \begin{pmatrix}
    0&0&0&t_{16}&t_{17}&t_{18}\\
    0&0&0&t_{17}&t_{19}&t_{20}\\
    0&0&0&t_{18}&t_{20}&t_{21}\\
    t_{16}&t_{17}&-t_{18}&0&0&0\\
    t_{17}&t_{19}&-t_{20}&0&0&0\\
    -t_{18}&-t_{20}&t_{21}&0&0&0\\
    \end{pmatrix}
\end{align}

There also exist restrictions on entries in the sub-blocks due to $C_2^{010}$ and Hermiticity. These are:

\begin{equation}
t_{18} = \sqrt{3}t_{17}
\end{equation}

\begin{equation}
t_{20} = \sqrt{3}(t_{21}-t_{19})
\end{equation}

%
%
%
%

Consequently there are only four independent parameters for $3^{rd}$ atomic neighbors. 

Finally , for $4^{th}$ atomic neighbor interactions, we look at the off-diagnoal terms of $H_{100}^{4}$:

\begin{align}
    H_{100}^{4} =
    \begin{pmatrix}
    0&0&0&t_{22}&t_{23}&t_{24}\\
    0&0&0&t_{23}&t_{25}&t_{26}\\
    0&0&0&t_{24}&t_{26}&t_{27}\\
    t_{22}&t_{23}&t_{24}&0&0&0\\
    t_{23}&t_{25}&t_{26}&0&0&0\\
    t_{24}&t_{26}&t_{27}&0&0&0\\
    \end{pmatrix}
\end{align}

There are six independent entries in the sub-block of the real space Hamiltonian $H_{100}^{4}$  due to inversion symmetry and Hermiticity. The rest of the hopping terms can be generated easily using the other $C_2$ and $C_3^{\pm}$ rotations and taking the Hermitian adjoint. Now we have finished construction of spinless Hamiltonian with up to $4^{th}$ neighbor interaction included using the form of real space Hamiltonian.

Nearest neighbors:
\begin{equation*}
H^{12}_{[010]} = D(C^+_3)H^{12}_{[000]}D^{-1}(C^+_3)
\end{equation*}

\begin{equation*}
H^{12}_{[\bar{1}00]} = D(C^-_3)H^{12}_{[000]}D^{-1}(C^-_3)
\end{equation*}

\begin{equation*}
H^{21}_{[000]} = H^{12 \dagger}_{[000]}
\end{equation*}

\begin{equation*}
H^{21}_{[0\bar{1}0]} = D(C^+_3)H^{21}_{[000]}D^{-1}(C^+_3)
\end{equation*}

\begin{equation*}
H^{21}_{[100]} = D(C^-_3)H^{21}_{[000]}D^{-1}(C^-_3)
\end{equation*}

$2^{nd}$ Nearest neighbors:

\begin{equation*}
H^{11}_{[010]} = D(C^+_3)H^{11}_{[100]}D^{-1}(C^+_3)
\end{equation*}

\begin{equation*}
H^{11}_{[\bar{1}\bar{1}0]} = D(C^-_3)H^{11}_{[100]}D^{-1}(C^-_3)
\end{equation*}

\begin{equation*}
H^{11}_{[110]} = D(\sigma_{010})H^{11}_{[100]}D^{-1}(\sigma_{010})
\end{equation*}

\begin{equation*}
H^{11}_{[\bar{1}00]} = D(\sigma_{100})H^{11}_{[100]}D^{-1}(\sigma_{100})
\end{equation*}

\begin{equation*}
H^{11}_{[0\bar{1}0]} = D(\sigma_{110})H^{11}_{[100]}D^{-1}(\sigma_{110})
\end{equation*}

\begin{equation*}
H^{22}_{[100]} = H^{11 \dagger}_{[100]}
\end{equation*}

\begin{equation*}
H^{22}_{[010]} = D(C^+_3)H^{22}_{[100]}D^{-1}(C^+_3)
\end{equation*}

\begin{equation*}
H^{22}_{[\bar{1}\bar{1}0]} = D(C^-_3)H^{22}_{[100]}D^{-1}(C^-_3)
\end{equation*}

\begin{equation*}
H^{22}_{[110]} = D(\sigma_{010})H^{22}_{[100]}D^{-1}(\sigma_{010})
\end{equation*}

\begin{equation*}
H^{22}_{[\bar{1}00]} = D(\sigma_{100})H^{22}_{[100]}D^{-1}(\sigma_{100})
\end{equation*}

\begin{equation*}
H^{22}_{[0\bar{1}0]} = D(\sigma_{110})H^{22}_{[100]}D^{-1}(\sigma_{110})
\end{equation*}

$3^{rd}$ Nearest neighbors:
\begin{equation*}
H^{21}_{[110]}=D(C_2^{[110]})H^{12}_{[110]}D^{-1}(C_2^{[110]})
\end{equation*}

\begin{equation*}
H^{12}_{[\bar{1}10]}=D(C_3^+)H^{12}_{[110]}D^{-1}(C_3^+)
\end{equation*}

\begin{equation*}
H^{12}_{[\bar{1}\bar{1}0]}=D(C_3^-)H^{12}_{[110]}D^{-1}(C_3^-)
\end{equation*}

\begin{equation*}
H^{21}_{[\bar{1}\bar{1}0]}=D(C_3^+)H^{21}_{[110]}D^{-1}(C_3^+)
\end{equation*}

\begin{equation*}
H^{21}_{[1\bar{1}0]}=D(C_3^-)H^{21}_{[110]}D^{-1}(C_3^-)
\end{equation*}

$4^{th}$ Nearest neighbor
\begin{equation*}
H^{21}_{[\bar{1}00]}=D(i)H^{12}_{[100]}D^{-1}(i)
\end{equation*}

\begin{equation*}
H^{12}_{[020]}=D(C_3^+)H^{12}_{[100]}D^{-1}(C_3^+)
\end{equation*}

\begin{equation*}
H^{21}_{[0\bar{2}0]}=D(S_6^-)H^{12}_{[100]}D^{-1}(S_6^-)
\end{equation*}

\begin{equation*}
H^{12}_{[\bar{2}\bar{1}0]}=D(C_3^-)H^{12}_{[100]}D^{-1}(S_6^-)
\end{equation*}

\begin{equation*}
H^{21}_{[210]}=D(S_6^+)H^{12}_{[100]}D^{-1}(S_6^+)
\end{equation*}

\begin{equation*}
H^{21}_{[010]}=D(C_2^{[110]})H^{12}_{[100]}D^{-1}(C_2^{[110]})
\end{equation*}

\begin{equation*}
H^{12}_{[0\bar{1}0]}=D(\sigma_{110})H^{12}_{[100]}D^{-1}(\sigma_{110})
\end{equation*}

\begin{equation*}
H^{21}_{[200]}=D(C_2^{[100]})H^{12}_{[100]}D^{-1}(C_2^{[100]})
\end{equation*}

\begin{equation*}
H^{12}_{[\bar{2}00]}=D(\sigma_{100})H^{12}_{[100]}D^{-1}(\sigma_{100})
\end{equation*}

\begin{equation*}
H^{21}_{[\bar{1}\bar{2}0]}=D(C_2^{[010]})H^{12}_{[100]}D^{-1}(C_2^{[010]})
\end{equation*}

\begin{equation*}
H^{12}_{[120]}=D(\sigma_{010})H^{12}_{[100]}D^{-1}(\sigma_{010})
\end{equation*}

\subsubsection{Symmetry constraints on SOC for the Bi (111) bilayer}

In the Bi (111) bilayer we consider two types of SOC terms, namely on-site SOC and a next-nearest atomic neighbor SOC referred to as intrinsic. The on-site SOC can be directly calculated using Eq. 23 and has the form:

\setcounter{MaxMatrixCols}{12}
\begin{align*}
    H_{soc} = 
    \begin{pmatrix}
    0&0&0&0&0&0&0&-1&i&0&0&0\\
    0&0&-i&0&0&0&1&0&0&0&0&0\\
    0&i&0&0&0&0&-i&0&0&0&0&0\\
    0&0&0&0&0&0&0&0&0&0&-1&i\\
    0&0&0&0&0&-i&0&0&0&1&0&0\\
    0&0&0&0&i&0&0&0&0&-i&0&0\\
    0&1&i&0&0&0&0&0&0&0&0&0\\
    -1&0&0&0&0&0&0&0&i&0&0&0\\
    -i&0&0&0&0&0&0&-i&0&0&0&0\\
    0&0&0&0&1&i&0&0&0&0&0&0\\
    0&0&0&-1&0&0&0&0&0&0&0&i\\
    0&0&0&-i&0&0&0&0&0&0&-i&0\\
    \end{pmatrix}
\end{align*}

Next we turn to the intrinsic SOC. The form of this SOC can be obtained by using the approach in the previous section for deriving hopping terms. The crystal symmetries, TR symmetry and Hermicity play important  roles in limiting the number of independent SOC terms. 
 
We can write the next-nearest atomic neighbor SOC hopping terms by investigating the diagonal terms of real space intrinsic SOC Hamiltonian $H_I[110]$:
\begin{equation}
H_I[110] = i\sigma_z \otimes 
\begin{pmatrix}
H_{I,11}[\bar{1}\bar{1}0]& 0\\
0 & H_{I,11}[\bar{1}\bar{1}0]^\dagger,
\end{pmatrix}
\end{equation}
where $\sigma_z$ is a Pauli matrix and $H_{I,11}[\bar{1}\bar{1}0]$ represents the SOC between atom 1 in the home cell and atom 1 in cell $\bm{R}=[\bar{1}\bar{1}0]$. $H_{I,11}[\bar{1}\bar{1}0]$ has the form:\\
\begin{equation*}
\begin{pmatrix}
\lambda_1 & \lambda_2 &\lambda_3\\
\lambda_2& \lambda_4 &\lambda_5\\
-\lambda_3&-\lambda_5&\lambda_6\\
\end{pmatrix}
\end{equation*}

Using the findings in the previous section, when the symmetry constraints act on the next-nearest hopping terms, there will only be six independent terms. We can therefore include the contribution of SOC in a Bi (111) bilayer through an on-site SOC strength $\lambda$ and next-nearest SOC strengths $\lambda_{1}$ to $\lambda_{6}$.

\subsubsection{Bismuthene}

In the case of bismuthene, we construct a real space Hamiltonian by including hopping terms to the next-nearest atomic neighbor. Similarly to a Bi (111) bilayer, we give the form of Hamiltonian for cell $\bm{R}=[000]$ as:

\begin{equation}
H_{[000]} =
\begin{pmatrix}
t_1&0&0&t_3&0&0\\
0&t_2&0&0&t_4&0\\
0&0&t_2&0&0&t_5\\
t_3&0&0&t_1&0&0\\
0&t_4&0&0&t_2&0\\
0&0&t_5&0&0&t_2\\
\end{pmatrix}
\end{equation}

The next-nearest neighbor hopping terms can be obtained by looking at diagonal sub-blocks of $H_{[100]}$:

\begin{equation}
H_{[100]} =
\begin{pmatrix}
t_6&0&0&0&0&0\\
0&t_7&t_8&0&0&0\\
0&t_9&t_{10}&0&0&0\\
0&0&0&t_6&0&0\\
0&0&0&0&t_7&t_8\\
0&0&0&0&t_9&t_{10}\\
\end{pmatrix}
\end{equation}

Applying the same symmetry operations of a (111) bilayer listed in the previous section on the real space Hamiltonian of Bismuthene, the real space Hamiltonian including up to NN interactions can be obtained. We describe the electronic structure of bismuthene with nine independent parameters as shown in Table \ref{prm_bismuthene}. The number of hopping terms are reduced by the presence of crystal symmetries and Hermicity. When considering  SOC, the form of on-site SOC Hamiltonian of bismuthene is identical to that of Bi (111) bilayer.


\subsubsection{Bi (110) bilayer}

For a Bi (110) bilayaer, we use a neighboring cell instead of the neigboring atoms, but the explicit form of the real space Hamiltonian of cell $\bm{R}=[uvw]$ still follows the definition as described for the Bi (111) bilayer and bismuthene. For cell $\bm{R} = [000]$, we write the sub-blocks representing the interactions between one lattice site in the home cell $\bm{R} = [000]$ and another lattice site in the home cell as, with superscript indicating the sub-lattice sites and subscript the cell that interacts with the home cell:

\begin{equation}
H^{11}_{[000]} = 
\begin{pmatrix}
t_1&0&t_4\\
0&t_2&0\\
t_4&0&t_{3}\\
\end{pmatrix}
\end{equation}
\begin{equation}
H^{12}_{[000]} = 
\begin{pmatrix}
t_{13}&t_{16}&t_{17}\\
t_{18}&t_{14}&t_{19}\\
0&0&t_{15}\\
\end{pmatrix}
\end{equation}
\begin{equation}
H^{13}_{[000]} = 
\begin{pmatrix}
t_{5}&0&t_{8}\\
0&t_{6}&0\\
t_8&0&t_{7}\\
\end{pmatrix}
\end{equation}
\begin{equation}
H^{14}_{[000]} = 
\begin{pmatrix}
t_{9}&0&0\\
0&t_{10}&-t_{12}\\
0&-t_{12}&t_{11}\\
\end{pmatrix}
\end{equation}

Then the rest of the hopping terms for cell $\bm{R} = [000]$ can be derived by applying symmetry operations on sub-blocks one by one. \\
\begin{equation*}
H^{21 \dagger}_{[000]}=H^{12}_{[000]} \\
\end{equation*}
\begin{equation*}
H^{22}_{[000]}=D(\sigma_{z})H^{11}_{[000]}D^{-1}(\sigma_{z})
\end{equation*}
\begin{equation*}
H^{23}_{[000]}=D(C_{2y}) H^{11}_{[000]}D^{-1}(C_{2y})
\end{equation*}
\begin{equation*}
H^{31}_{[000]}=H^{13 \dagger}_{[000]}
\end{equation*}
\begin{equation*}
H^{32}_{[000]} = [D(C_{2y})H^{11}_{[000]}D^{-1}(C_{2y})]^\dagger
\end{equation*}
\begin{equation*}
H^{33}_{[000]} = D(i)H^{11}_{[000]} D^{-1}(i)
\end{equation*}
\begin{equation*}
H^{34}_{[000]} = D(C_{2x})H^{12}_{[000]} D^{-1}(C_{2x})
\end{equation*}
\begin{equation*}
H^{41}_{[000]} = H^{14 \dagger}_{[000]} 
\end{equation*}
\begin{equation*}
H^{43}_{[000]} = H^{34 \dagger}_{[000]} 
\end{equation*}
\begin{equation*}
H^{44}_{[000]} = D(C_{2z})H^{12}_{[000]} D^{-1}(C_{2z})
\end{equation*}

For hopping terms from lattice sites in the home cell to surrounding cells, we start by considering hopping to cell $\bm{R}=[100]$. The only two independent sub-blocks are $H^{11}_{[100]}$ and $H^{24}_{[100]}$:

\begin{equation}
H^{11}_{[100]} = 
\begin{pmatrix}
0&0&0\\
0&0&t_{21}\\
0&-t_{21}&t_{20}\\
\end{pmatrix}
\end{equation}
\begin{equation}
H^{24}_{[100]} = 
\begin{pmatrix}
0&0&0\\
0&t_{22}&0\\
0&0&t_{23}\\
\end{pmatrix}
\end{equation}

\begin{equation*}
H^{21}_{[100]} = D(\sigma_x)H^{12 \dagger}_{[000]}D^{-1}(\sigma_x)
\end{equation*}
\begin{equation*}
H^{22}_{[100]} = D(\sigma_z)H^{11}_{[100]}D^{-1}(\sigma_z)
\end{equation*}
\begin{equation*}
H^{23}_{[100]} = D(\sigma_x)H^{23}_{[000]}D^{-1}(\sigma_x)
\end{equation*}
\begin{equation*}
H^{33}_{[100]} = D(C_{2x})H^{12 \dagger}_{[100]}D^{-1}(C_{2x})
\end{equation*}
\begin{equation*}
H^{41}_{[100]} = D(\sigma_y)H^{14}_{[000]}D^{-1}(\sigma_y)
\end{equation*}
\begin{equation*}
H^{42}_{[100]} = D(C_{2x})H^{24}_{[100]}D^{-1}(C_{2x})
\end{equation*}
\begin{equation*}
H^{43}_{[100]} = D(i)H^{21}_{[000]}D^{-1}(i)
\end{equation*}
\begin{equation*}
H^{44}_{[100]} = D(\sigma_y)H^{11 \dagger}_{[100]}D^{-1}(\sigma_y)
\end{equation*}

For hopping to cell $\bm{R}=[010]$, the two independent sub-blocks are:

\begin{equation}
H^{11}_{[010]} = 
\begin{pmatrix}
0&0&0\\
0&0&0\\
0&0&t_{24}\\
\end{pmatrix}
\end{equation}
\begin{equation}
H^{14}_{[010]} = 
\begin{pmatrix}
t_{25}&t_{28}&t_{29}\\
-t_{28}&t_{26}&t_{30}\\
-t_{29}&t_{30}&t_{27}\\
\end{pmatrix}
\end{equation}

\begin{equation*}
H^{21}_{[010]} = D(C_{2y})H^{12}_{[000]}D^{-1}(C_{2y})
\end{equation*}
\begin{equation*}
H^{22}_{[010]} = D(C_{2y})H^{11}_{[010]}D^{-1}(C_{2y})
\end{equation*}
\begin{equation*}
H^{23}_{[010]} = D(C_{2y})H^{14}_{[010]}D^{-1}(C_{2y})
\end{equation*}
\begin{equation*}
H^{24}_{[010]} = D(C_{2y})H^{13}_{[000]}D^{-1}(C_{2y})
\end{equation*}
\begin{equation*}
H^{33}_{[010]} = D(i)H^{11 \dagger}_{[010]}D^{-1}(i)
\end{equation*}
\begin{equation*}
H^{34}_{[010]} = D(C_{2z})H^{21}_{[000]}D^{-1}(C_{2z})
\end{equation*}
\begin{equation*}
H^{44}_{[010]} = D(\sigma_z)H^{33}_{[010]}D^{-1}(\sigma_z)
\end{equation*}

For hopping to cell $\bm{R}=[110]$, the only independent sub-block is:

\begin{equation}
H^{11}_{[110]} = 
\begin{pmatrix}
t_{31}&t_{34}&t_{35}\\
-t_{34}&t_{32}&-t_{23}\\
t_{35}&t_{23}&t_{33}\\
\end{pmatrix}
\end{equation}

\begin{equation*}
H^{21}_{[110]} = D(\sigma_z)H^{12}_{[000]}D^{-1}(\sigma_z)
\end{equation*}
\begin{equation*}
H^{22}_{[110]} = D(\sigma_z)H^{11}_{[110]}D^{-1}(\sigma_z)
\end{equation*}

\begin{equation*}
H^{23}_{[110]} = D(\sigma_x)H^{23}_{[010]}D^{-1}(\sigma_x)
\end{equation*}
\begin{equation*}
H^{31}_{[110]} = D(\sigma_y)H^{24}_{[100]}D^{-1}(\sigma_y)
\end{equation*}
\begin{equation*}
H^{33}_{[110]} = D(i)H^{11 \dagger}_{[110]}D^{-1}(i)
\end{equation*}
\begin{equation*}
H^{44}_{[110]} = D(\sigma_z)H^{33}_{[110]}D^{-1}(\sigma_z)
\end{equation*}

For hopping to cell $\bm{R} = [1\bar{1}0]$, all the sub-blocks can be generated from the sub-blocks above:
\begin{equation*}
H^{11}_{[1\bar{1}0]} = D(C_{2x})H^{33}_{[110]}D^{-1}(C_{2x})
\end{equation*}
\begin{equation*}
H^{13}_{[1\bar{1}0]} = D(C_{2x})H^{31}_{[110]}D^{-1}(C_{2x})
\end{equation*}
\begin{equation*}
H^{22}_{[1\bar{1}0]} = D(C_{2x})H^{44}_{[110]}D^{-1}(C_{2x})
\end{equation*}
\begin{equation*}
H^{33}_{[1\bar{1}0]} = D(C_{2x})H^{11}_{[110]}D^{-1}(C_{2x})
\end{equation*}
\begin{equation*}
H^{41}_{[1\bar{1}0]} = D(C_{2x})H^{23}_{[110]}D^{-1}(C_{2x})
\end{equation*}
\begin{equation*}
H^{43}_{[1\bar{1}0]} = D(C_{2x})H^{21}_{[110]}D^{-1}(C_{2x})
\end{equation*}
\begin{equation*}
H^{44}_{[1\bar{1}0]} = D(C_{2x})H^{22}_{[110]}D^{-1}(C_{2x})
\end{equation*}

Lastly, for hopping to cell $\bm{R}=[\bar{1}00], \bm{R}=[0\bar{1}0], \bm{R}=[\bar{1}\bar{1}0]$ and $\bm{R}=[\bar{1}10]$, they can be obtained by taking Hermit conjugate of the hopping listed above:
\begin{equation*}
H_[\bar{1}00]=H^\dagger_{[100]}
\end{equation*}
\begin{equation*}
H_[0\bar{1}0]=H^\dagger_{[010]}
\end{equation*}
\begin{equation*}
H_[\bar{1}\bar{1}0]=H^\dagger_{[110]}
\end{equation*}
\begin{equation*}
H_[\bar{1}10]=H^\dagger_{[1\bar{1}0]}
\end{equation*}

\subsection{Character table for symmetry groups of three allotropes}

The character tables for a Bi (111) bilayer and bismuthene can be derived from their corresponding point groups. Those character tables can be readily found in Koster's book \cite{koster}. Here we show the character table for a Bi (111) bilayer ($D_{3d}$) and bismuthene ($D_{6h}$) in Table \ref{D3d} and Table \ref{D6h}, respectively. The character table for a Bi (110) bilayer involves nonsymmporhic symmetries and the characters for $\Gamma$, $M$, $X$ and $Y$ point are listed in Table \ref{D2h} to Table \ref{Y}.
\begin{table*}[hp]
    \caption{$D_{3d}$ Double Group Character Table.}
    \label{D3d}
    \begin{tabular*}{1.1\columnwidth}{@{\extracolsep{\fill}}c c c c c c c c c c c c c }
        \hline
        \hline
        $D_{3d}$ & E & $\bar{E}$ & $2C_3$ &$2\bar{C_3}$&$3C_2'$ &$3\bar{C_2'}$& $I$ & $\bar{I}$ &$2S_6$& $2\bar{S_6}$& 3$\sigma_d$ & 3$\bar{\sigma_d}$\\
        \hline
	$\Gamma_1^+$&1&1&1&1&1&1&1&1&1&1&1&1\\
	$\Gamma_2^+$&1&1&1&1&-1&-1&1&1&1&1&-1&-1\\
	$\Gamma_3^+$&2&2&-1&-1&0&0&2&2&-1&-1&0&0\\
	$\Gamma_1^-$&1&1&1&1&1&1&-1&-1&-1&-1&-1&-1\\
	$\Gamma_2^-$&1&1&1&1&-1&-1&-1&-1&-1&-1&1&1\\
	$\Gamma_3^-$&2&2&-1&-1&0&0&-2&-2&1&1&0&0\\
	\hline
	$\Gamma_4^+$&1&-1&-1&1&i&-i&1&-1&-1&1&i&-i\\
	$\Gamma_5^+$&1&-1&-1&1&-i&i&1&-1&-1&1&-i&i\\
	$\Gamma_6^+$&2&-2&1&-1&0&0&2&-2&1&-1&0&0\\
	$\Gamma_4^-$&1&-1&-1&1&i&-i&-1&1&1&-1&-i&i\\
	$\Gamma_5^-$&1&-1&-1&1&-i&i&-1&1&1&-1&i&-i\\
	$\Gamma_6^-$&2&-2&1&-1&0&0&-2&2&-1&1&0&0\\
        \hline
        \hline
    \end{tabular*}
\end{table*}

\begin{table*}[hp]
    \caption{$D_{6h}$ Double Group Character Table Part 1.}
    \label{D6h}
    \begin{tabular*}{1.1\columnwidth}{@{\extracolsep{\fill}}c c c c c c c c c  c }
        \hline
        \hline
        $D_{6h}$ & E & $\bar{E}$ & $\{C_2, \bar{C_2}\}$ & $2C_3$ & $2\bar{C_3}$&$2C_6$ &$2\bar{C_6}$& $\{3C_2', 3\bar{C_2'}\}$ & $\{3C_2'', 3\bar{C_2''}\}$\\
        \hline
	$\Gamma_1^+$&1&1&1&1&1&1&1&1&1\\
	$\Gamma_2^+$&1&1&1&1&1&1&1&-1&-1 \\
	$\Gamma_3^+$&1&1&-1&1&1&-1&-1&1&-1 \\
	$\Gamma_4^+$&1&1&-1&1&1&-1&-1&-1&1 \\
	$\Gamma_5^+$&2&2&-2&-1&-1&1&1&0&0 \\
	$\Gamma_6^+$&2&2&2&-1&-1&-1&-1&0&0 \\
	$\Gamma_1^-$&1&1&1&1&1&1&1&1&1\\
	$\Gamma_2^-$&1&1&1&1&1&1&1&-1&-1 \\
	$\Gamma_3^-$&1&1&-1&1&1&-1&-1&1&-1 \\
	$\Gamma_4^-$&1&1&-1&1&1&-1&-1&-1&1 \\
	$\Gamma_5^-$&2&2&-2&-1&-1&1&1&0&0 \\
	$\Gamma_6^-$&2&2&2&-1&-1&-1&-1&0&0 \\
	\hline
	$\Gamma_7^+$&2&-2&0&1&-1&$\sqrt{3}$&-$\sqrt{3}$&0&0 \\
	$\Gamma_8^+$&2&-2&0&1&-1&$-\sqrt{3}$&$\sqrt{3}$&0&0 \\
	$\Gamma_9^+$&2&-2&0&-2&2&0&0&0&0 \\
	$\Gamma_7^-$&2&-2&0&1&-1&$\sqrt{3}$&-$\sqrt{3}$&0&0 \\
	$\Gamma_8^-$&2&-2&0&1&-1&$-\sqrt{3}$&$\sqrt{3}$&0&0 \\
	$\Gamma_9^-$&2&-2&0&-2&2&0&0&0&0 \\
        \hline
        \hline

    \end{tabular*}
\end{table*}

\begin{table*}[hp]
    \caption{$D_{6h}$ Double Group Character Table Part 2.}
    \label{D6h}
    \begin{tabular*}{1.1\columnwidth}{@{\extracolsep{\fill}}c c c c c c c c c  c }
        \hline
        \hline
        $D_{6h}$ & I & $\bar{I}$ & $\{\sigma_h, \bar{\sigma_h}\}$ & $2S_6$ & $2\bar{S_6}$&$2S_3$ &$2\bar{S_3}$& $\{3\sigma_d$, 3$\bar{\sigma_d}\}$ & $\{3\sigma_v$, 3$\bar{\sigma_v}\}$\\
        \hline
	$\Gamma_1^+$&1&1&1&1&1&1&1&1&1\\
	$\Gamma_2^+$&1&1&1&1&1&1&1&-1&-1 \\
	$\Gamma_3^+$&1&1&-1&1&1&-1&-1&1&-1 \\
	$\Gamma_4^+$&1&1&-1&1&1&-1&-1&-1&1 \\
	$\Gamma_5^+$&2&2&-2&-1&-1&1&1&0&0 \\
	$\Gamma_6^+$&2&2&2&-1&-1&-1&-1&0&0 \\
	$\Gamma_1^-$&-1&-1&-1&-1&-1&-1&-1&-1&-1\\
	$\Gamma_2^-$&-1&-1&-1&-1&-1&-1&-1&1&1 \\
	$\Gamma_3^-$&-1&-1&1&-1&-1&1&1&-1&1 \\
	$\Gamma_4^-$&-1&-1&1&-1&-1&1&1&1&-1 \\
	$\Gamma_5^-$&-2&-2&2&1&1&-1&-1&0&0 \\
	$\Gamma_6^-$&-2&-2&-2&1&1&1&1&0&0 \\
	\hline
	$\Gamma_7^+$&2&-2&0&1&-1&$\sqrt{3}$&-$\sqrt{3}$&0&0 \\
	$\Gamma_8^+$&2&-2&0&1&-1&$-\sqrt{3}$&$\sqrt{3}$&0&0 \\
	$\Gamma_9^+$&2&-2&0&-2&2&0&0&0&0 \\
	$\Gamma_7^-$&-2&2&0&-1&1&$-\sqrt{3}$&$\sqrt{3}$&0&0 \\
	$\Gamma_8^-$&-2&2&0&-1&1&$\sqrt{3}$&$-\sqrt{3}$&0&0 \\
	$\Gamma_9^-$&-2&2&0&2&-2&0&0&0&0 \\
        \hline
        \hline

    \end{tabular*}
\end{table*}

\begin{table*}[htbp]
    \caption{$D_{2h}$ Double Space Group Character Table $u=(\frac{1}{2},\frac{1}{2},0)$}
    \label{D2h}
    \begin{tabular*}{1.9\columnwidth}{@{\extracolsep{\fill}}c c c c c c c c c c c c }
        \hline
        \hline
        $D_{2h}$ & E & $\bar{E}$ & $\{C_{2x}, \bar{C_{2x}} \}$ & $\{\{C_{2z}|u\}, \{\bar{C_{2z}}|u\} \}$ &$\{\{C_{2y}|u\}, \{\bar{C_{2y}}|u\} \}$ &$I$& $\bar{I}$ & $\{\sigma_x, \bar{\sigma_x}\}$ &$\{\{\sigma_z |u\}, \{\bar{\sigma_z}|u\}\}$ & $\{\{\sigma_y |u\}, \{\bar{\sigma_y}|u\}\}$ \\
        \hline
	$\Gamma_1^+$&1&1&1&1&1&1&1&1&1&1\\
	$\Gamma_2^+$&1&1&-1&1&-1&1&1&-1&1&-1\\
	$\Gamma_3^+$&1&1&1&-1&-1&1&1&1&-1&-1\\
	$\Gamma_4^+$&1&1&-1&-1&1&1&1&-1&-1&1\\
	$\Gamma_1^-$&1&1&1&1&1&-1&-1&-1&-1&-1\\
	$\Gamma_2^-$&1&1&-1&1&-1&-1&-1&1&-1&1\\
	$\Gamma_3^-$&1&1&1&-1&-1&-1&-1&-1&1&1\\
	$\Gamma_4^-$&1&1&-1&-1&1&-1&-1&1&1&-1\\
	\hline
	$\Gamma_5^+$&2&-2&0&0&0&2&-2&0&0&0\\
	$\Gamma_5^-$&2&-2&0&0&0&-2&2&0&0&0\\

        \hline
    \end{tabular*}
\end{table*}

\begin{table*}[!htbp]
    \caption{Double Little Group Character Table at M point in FBZ of Bi (110) $u=(\frac{1}{2},\frac{1}{2},0)$}
    \label{M}
    \begin{tabular*}{1.9\columnwidth}{@{\extracolsep{\fill}}c c c c c c c c c c c c c c c c c }
        \hline
        \hline
        $M$ & E & $\{C_{2y}|u\} $ & $\{C_{2z}|u\}$&$C_{2x}$ & $I$ &$\{\sigma_y |u\}$ &$\{\sigma_z |u\}$& $\sigma_x$ &$\bar{E}$&$ \{\bar{C_{2y}}|u\}$& $\{\bar{C_{2z}}|u\}$& $\bar{C_{2x}}$ & $\bar{I}$& $\{\bar{\sigma_y}|u\}$& $\{\bar{\sigma_z}|u\}$& $\bar{\sigma_x}$\\
        \hline
	$M_1^+$&2&0&0&0&2&0&0&0&2&0&0&0&2&0&0&0\\
	$M_1^-$&2&0&0&0&-2&0&0&0&2&0&0&0&-2&0&0&0\\
	\hline
	$M_2^+$&1&1&-i&-i&1&1&-i&-i&-1&-1&i&i&-1&-1&i&i\\
	$M_3^+$&1&-1&i&-i&1&-1&i&-i&-1&1&-i&i&-1&1&-i&i\\
	$M_4^+$&1&1&i&i&1&1&i&i&-1&-1&-i&-i&-1&-1&-i&-i\\
	$M_5^+$&1&-1&-i&i&1&-1&-i&i&-1&1&i&-i&-1&1&i&-i\\
	$M_2^-$&1&1&-i&-i&-1&-1&i&i&-1&-1&i&i&1&1&-i&-i\\
	$M_3^-$&1&-1&i&-i&-1&1&-i&i&-1&1&-i&i&1&-1&i&-i\\
	$M_4^-$&1&1&i&i&-1&-1&-i&-i&-1&-1&-i&-i&1&1&i&i\\
	$M_5^-$&1&-1&-i&i&-1&1&i&-i&-1&1&i&-i&1&-1&-i&i\\
        \hline
    \end{tabular*}
\end{table*}

\begin{table*}[!htbp]
    \caption{Double Little Group Character Table at X point in FBZ of Bi (110) $u=(\frac{1}{2},\frac{1}{2},0)$}
    \label{X}
    \begin{tabular*}{1.9\columnwidth}{@{\extracolsep{\fill}}c c c c c c c c c c c c c c c c c }
        \hline
        \hline
        $X$ & E & $\{C_{2y}|u\} $ & $\{C_{2z}|u\}$&$C_{2x}$ & $I$ &$\{\sigma_y |u\}$ &$\{\sigma_z |u\}$& $\sigma_x$ &$\bar{E}$&$ \{\bar{C_{2y}}|u\}$& $\{\bar{C_{2z}}|u\}$& $\bar{C_{2x}}$ & $\bar{I}$& $\{\bar{\sigma_y}|u\}$& $\{\bar{\sigma_z}|u\}$& $\bar{\sigma_x}$\\
        \hline
	$X_1$&2&0&0&2&0&0&0&0&2&0&0&2&0&0&0&0\\
	$X_2$&2&0&0&-2&0&0&0&0&2&0&0&-2&0&0&0&0\\
	\hline
	$X_3$&2&0&0&0&0&0&0&-2i&-2&0&0&0&0&0&0&2i\\
	$X_4$&2&0&0&0&0&0&0&2i&-2&0&0&0&0&0&0&-2i\\
        \hline
    \end{tabular*}
\end{table*}

\begin{table*}[!htbp]
    \caption{Double Little Group Character Table at Y point in FBZ of Bi (110) $u=(\frac{1}{2},\frac{1}{2},0)$}
    \label{Y}
    \begin{tabular*}{1.9\columnwidth}{@{\extracolsep{\fill}}c c c c c c c c c c c c c c c c c }
        \hline
        \hline
        $Y$ & E & $\{C_{2y}|u\} $ & $\{C_{2z}|u\}$&$C_{2x}$ & $I$ &$\{\sigma_y |u\}$ &$\{\sigma_z |u\}$& $\sigma_x$ &$\bar{E}$&$ \{\bar{C_{2y}}|u\}$& $\{\bar{C_{2z}}|u\}$& $\bar{C_{2x}}$ & $\bar{I}$& $\{\bar{\sigma_y}|u\}$& $\{\bar{\sigma_z}|u\}$& $\bar{\sigma_x}$\\
        \hline
	$Y_1$&2&0&0&0&0&0&0&2&2&0&0&0&0&0&0&2\\
	$Y_2$&2&0&0&0&0&0&0&-2&2&0&0&0&0&0&0&-2\\
	\hline
	$Y_3$&2&0&0&-2i&0&0&0&0&-2&0&0&2i&0&0&0&0\\
	$Y_4$&2&0&0&2i&0&0&0&0&-2&0&0&-2i&0&0&0&0\\
        \hline
    \end{tabular*}
\end{table*}


\begin{thebibliography}{46}%
\makeatletter
\providecommand \@ifxundefined [1]{%
 \@ifx{#1\undefined}
}%
\providecommand \@ifnum [1]{%
 \ifnum #1\expandafter \@firstoftwo
 \else \expandafter \@secondoftwo
 \fi
}%
\providecommand \@ifx [1]{%
 \ifx #1\expandafter \@firstoftwo
 \else \expandafter \@secondoftwo
 \fi
}%
\providecommand \natexlab [1]{#1}%
\providecommand \enquote  [1]{``#1''}%
\providecommand \bibnamefont  [1]{#1}%
\providecommand \bibfnamefont [1]{#1}%
\providecommand \citenamefont [1]{#1}%
\providecommand \href@noop [0]{\@secondoftwo}%
\providecommand \href [0]{\begingroup \@sanitize@url \@href}%
\providecommand \@href[1]{\@@startlink{#1}\@@href}%
\providecommand \@@href[1]{\endgroup#1\@@endlink}%
\providecommand \@sanitize@url [0]{\catcode `\\12\catcode `\$12\catcode
  `\&12\catcode `\#12\catcode `\^12\catcode `\_12\catcode `\%12\relax}%
\providecommand \@@startlink[1]{}%
\providecommand \@@endlink[0]{}%
\providecommand \url  [0]{\begingroup\@sanitize@url \@url }%
\providecommand \@url [1]{\endgroup\@href {#1}{\urlprefix }}%
\providecommand \urlprefix  [0]{URL }%
\providecommand \Eprint [0]{\href }%
\providecommand \doibase [0]{http://dx.doi.org/}%
\providecommand \selectlanguage [0]{\@gobble}%
\providecommand \bibinfo  [0]{\@secondoftwo}%
\providecommand \bibfield  [0]{\@secondoftwo}%
\providecommand \translation [1]{[#1]}%
\providecommand \BibitemOpen [0]{}%
\providecommand \bibitemStop [0]{}%
\providecommand \bibitemNoStop [0]{.\EOS\space}%
\providecommand \EOS [0]{\spacefactor3000\relax}%
\providecommand \BibitemShut  [1]{\csname bibitem#1\endcsname}%
\let\auto@bib@innerbib\@empty
\bibitem [{\citenamefont {Hasan}\ and\ \citenamefont
  {Kane}(2010)}]{RevModPhys.82.3045}%
  \BibitemOpen
  \bibfield  {author} {\bibinfo {author} {\bibfnamefont {M.~Z.}\ \bibnamefont
  {Hasan}}\ and\ \bibinfo {author} {\bibfnamefont {C.~L.}\ \bibnamefont
  {Kane}},\ }\href {\doibase 10.1103/RevModPhys.82.3045} {\bibfield  {journal}
  {\bibinfo  {journal} {Rev. Mod. Phys.}\ }\textbf {\bibinfo {volume} {82}},\
  \bibinfo {pages} {3045} (\bibinfo {year} {2010})}\BibitemShut {NoStop}%
\bibitem [{\citenamefont {Bansil}\ \emph {et~al.}(2016)\citenamefont {Bansil},
  \citenamefont {Lin},\ and\ \citenamefont {Das}}]{RevModPhys.88.021004}%
  \BibitemOpen
  \bibfield  {author} {\bibinfo {author} {\bibfnamefont {A.}~\bibnamefont
  {Bansil}}, \bibinfo {author} {\bibfnamefont {H.}~\bibnamefont {Lin}}, \ and\
  \bibinfo {author} {\bibfnamefont {T.}~\bibnamefont {Das}},\ }\href {\doibase
  10.1103/RevModPhys.88.021004} {\bibfield  {journal} {\bibinfo  {journal}
  {Rev. Mod. Phys.}\ }\textbf {\bibinfo {volume} {88}},\ \bibinfo {pages}
  {021004} (\bibinfo {year} {2016})}\BibitemShut {NoStop}%
\bibitem [{\citenamefont {Drozdov}\ \emph {et~al.}(2014)\citenamefont
  {Drozdov}, \citenamefont {Alexandradinata}, \citenamefont {Jeon},
  \citenamefont {Nadj-Perge}, \citenamefont {Ji}, \citenamefont {Cava},
  \citenamefont {Bernevig},\ and\ \citenamefont {Yazdani}}]{drozdov2014one}%
  \BibitemOpen
  \bibfield  {author} {\bibinfo {author} {\bibfnamefont {I.~K.}\ \bibnamefont
  {Drozdov}}, \bibinfo {author} {\bibfnamefont {A.}~\bibnamefont
  {Alexandradinata}}, \bibinfo {author} {\bibfnamefont {S.}~\bibnamefont
  {Jeon}}, \bibinfo {author} {\bibfnamefont {S.}~\bibnamefont {Nadj-Perge}},
  \bibinfo {author} {\bibfnamefont {H.}~\bibnamefont {Ji}}, \bibinfo {author}
  {\bibfnamefont {R.}~\bibnamefont {Cava}}, \bibinfo {author} {\bibfnamefont
  {B.~A.}\ \bibnamefont {Bernevig}}, \ and\ \bibinfo {author} {\bibfnamefont
  {A.}~\bibnamefont {Yazdani}},\ }\href@noop {} {\bibfield  {journal} {\bibinfo
   {journal} {Nature Phys.}\ }\textbf {\bibinfo {volume} {10}},\ \bibinfo
  {pages} {664} (\bibinfo {year} {2014})}\BibitemShut {NoStop}%
\bibitem [{\citenamefont {Liu}\ and\ \citenamefont {Allen}(1995)}]{Liu1995a}%
  \BibitemOpen
  \bibfield  {author} {\bibinfo {author} {\bibfnamefont {Y.}~\bibnamefont
  {Liu}}\ and\ \bibinfo {author} {\bibfnamefont {R.~E.}\ \bibnamefont
  {Allen}},\ }\href {\doibase 10.1103/PhysRevB.52.1566} {\bibfield  {journal}
  {\bibinfo  {journal} {Phys. Rev. B}\ }\textbf {\bibinfo {volume} {52}},\
  \bibinfo {pages} {1566} (\bibinfo {year} {1995})},\ \Eprint
  {http://arxiv.org/abs/arXiv:1107.0075v1} {arXiv:arXiv:1107.0075v1}
  \BibitemShut {NoStop}%
\bibitem [{\citenamefont {Liu}\ \emph {et~al.}(2011)\citenamefont {Liu},
  \citenamefont {Liu}, \citenamefont {Wu}, \citenamefont {Duan}, \citenamefont
  {Liu},\ and\ \citenamefont {Wu}}]{PhysRevLett.107.136805}%
  \BibitemOpen
  \bibfield  {author} {\bibinfo {author} {\bibfnamefont {Z.}~\bibnamefont
  {Liu}}, \bibinfo {author} {\bibfnamefont {C.-X.}\ \bibnamefont {Liu}},
  \bibinfo {author} {\bibfnamefont {Y.-S.}\ \bibnamefont {Wu}}, \bibinfo
  {author} {\bibfnamefont {W.-H.}\ \bibnamefont {Duan}}, \bibinfo {author}
  {\bibfnamefont {F.}~\bibnamefont {Liu}}, \ and\ \bibinfo {author}
  {\bibfnamefont {J.}~\bibnamefont {Wu}},\ }\href {\doibase
  10.1103/PhysRevLett.107.136805} {\bibfield  {journal} {\bibinfo  {journal}
  {Phys. Rev. Lett.}\ }\textbf {\bibinfo {volume} {107}},\ \bibinfo {pages}
  {136805} (\bibinfo {year} {2011})}\BibitemShut {NoStop}%
\bibitem [{\citenamefont {Murakami}(2006)}]{Murakami2006a}%
  \BibitemOpen
  \bibfield  {author} {\bibinfo {author} {\bibfnamefont {S.}~\bibnamefont
  {Murakami}},\ }\href {\doibase 10.1103/PhysRevLett.97.236805} {\bibfield
  {journal} {\bibinfo  {journal} {Phys. Rev. Lett.}\ }\textbf {\bibinfo
  {volume} {97}},\ \bibinfo {pages} {236805} (\bibinfo {year} {2006})},\
  \Eprint {http://arxiv.org/abs/0607001v2} {arXiv:0607001v2 [arXiv:cond-mat]}
  \BibitemShut {NoStop}%
\bibitem [{\citenamefont {Munoz}\ \emph {et~al.}(2016)\citenamefont {Munoz},
  \citenamefont {Vergniory}, \citenamefont {Rauch}, \citenamefont {Henk},
  \citenamefont {Chulkov}, \citenamefont {Mertig}, \citenamefont {Botti},
  \citenamefont {Marques},\ and\ \citenamefont
  {Romero}}]{munoz2016topological}%
  \BibitemOpen
  \bibfield  {author} {\bibinfo {author} {\bibfnamefont {F.}~\bibnamefont
  {Munoz}}, \bibinfo {author} {\bibfnamefont {M.}~\bibnamefont {Vergniory}},
  \bibinfo {author} {\bibfnamefont {T.}~\bibnamefont {Rauch}}, \bibinfo
  {author} {\bibfnamefont {J.}~\bibnamefont {Henk}}, \bibinfo {author}
  {\bibfnamefont {E.~V.}\ \bibnamefont {Chulkov}}, \bibinfo {author}
  {\bibfnamefont {I.}~\bibnamefont {Mertig}}, \bibinfo {author} {\bibfnamefont
  {S.}~\bibnamefont {Botti}}, \bibinfo {author} {\bibfnamefont {M.~A.}\
  \bibnamefont {Marques}}, \ and\ \bibinfo {author} {\bibfnamefont
  {A.}~\bibnamefont {Romero}},\ }\href@noop {} {\bibfield  {journal} {\bibinfo
  {journal} {Sci. Rep.}\ }\textbf {\bibinfo {volume} {6}},\ \bibinfo {pages}
  {21790} (\bibinfo {year} {2016})}\BibitemShut {NoStop}%
\bibitem [{\citenamefont {Hsu}\ \emph {et~al.}(2016)\citenamefont {Hsu},
  \citenamefont {Huang}, \citenamefont {Crisostomo}, \citenamefont {Yao},
  \citenamefont {Chuang}, \citenamefont {Liu}, \citenamefont {Wang},
  \citenamefont {Hsu}, \citenamefont {Lee}, \citenamefont {Lin},\ and\
  \citenamefont {Bansil}}]{Hsu2016a}%
  \BibitemOpen
  \bibfield  {author} {\bibinfo {author} {\bibfnamefont {C.-H.}\ \bibnamefont
  {Hsu}}, \bibinfo {author} {\bibfnamefont {Z.-Q.}\ \bibnamefont {Huang}},
  \bibinfo {author} {\bibfnamefont {C.~P.}\ \bibnamefont {Crisostomo}},
  \bibinfo {author} {\bibfnamefont {L.-Z.}\ \bibnamefont {Yao}}, \bibinfo
  {author} {\bibfnamefont {F.-C.}\ \bibnamefont {Chuang}}, \bibinfo {author}
  {\bibfnamefont {Y.-T.}\ \bibnamefont {Liu}}, \bibinfo {author} {\bibfnamefont
  {B.}~\bibnamefont {Wang}}, \bibinfo {author} {\bibfnamefont {C.-H.}\
  \bibnamefont {Hsu}}, \bibinfo {author} {\bibfnamefont {C.-C.}\ \bibnamefont
  {Lee}}, \bibinfo {author} {\bibfnamefont {H.}~\bibnamefont {Lin}}, \ and\
  \bibinfo {author} {\bibfnamefont {A.}~\bibnamefont {Bansil}},\ }\href
  {\doibase 10.1038/srep18993} {\bibfield  {journal} {\bibinfo  {journal} {Sci.
  Rep.}\ }\textbf {\bibinfo {volume} {6}},\ \bibinfo {pages} {18993} (\bibinfo
  {year} {2016})}\BibitemShut {NoStop}%
\bibitem [{\citenamefont {Reis}\ \emph {et~al.}(2017)\citenamefont {Reis},
  \citenamefont {Li}, \citenamefont {Dudy}, \citenamefont {Bauernfeind},
  \citenamefont {Glass}, \citenamefont {Hanke}, \citenamefont {Thomale},
  \citenamefont {Sch{\"{a}}fer},\ and\ \citenamefont {Claessen}}]{Reis2017a}%
  \BibitemOpen
  \bibfield  {author} {\bibinfo {author} {\bibfnamefont {F.}~\bibnamefont
  {Reis}}, \bibinfo {author} {\bibfnamefont {G.}~\bibnamefont {Li}}, \bibinfo
  {author} {\bibfnamefont {L.}~\bibnamefont {Dudy}}, \bibinfo {author}
  {\bibfnamefont {M.}~\bibnamefont {Bauernfeind}}, \bibinfo {author}
  {\bibfnamefont {S.}~\bibnamefont {Glass}}, \bibinfo {author} {\bibfnamefont
  {W.}~\bibnamefont {Hanke}}, \bibinfo {author} {\bibfnamefont
  {R.}~\bibnamefont {Thomale}}, \bibinfo {author} {\bibfnamefont
  {J.}~\bibnamefont {Sch{\"{a}}fer}}, \ and\ \bibinfo {author} {\bibfnamefont
  {R.}~\bibnamefont {Claessen}},\ }\href {\doibase 10.1126/science.aai8142}
  {\bibfield  {journal} {\bibinfo  {journal} {Science}\ }\textbf {\bibinfo
  {volume} {357}},\ \bibinfo {pages} {287} (\bibinfo {year}
  {2017})}\BibitemShut {NoStop}%
\bibitem [{\citenamefont {Schindler}\ \emph {et~al.}(2018)\citenamefont
  {Schindler}, \citenamefont {Wang}, \citenamefont {Vergniory}, \citenamefont
  {Cook}, \citenamefont {Murani}, \citenamefont {Sengupta}, \citenamefont
  {Kasumov}, \citenamefont {Deblock}, \citenamefont {Jeon}, \citenamefont
  {Drozdov}, \citenamefont {Bouchiat}, \citenamefont {Gu{\'{e}}ron},
  \citenamefont {Yazdani}, \citenamefont {Bernevig},\ and\ \citenamefont
  {Neupert}}]{Schindler2018a}%
  \BibitemOpen
  \bibfield  {author} {\bibinfo {author} {\bibfnamefont {F.}~\bibnamefont
  {Schindler}}, \bibinfo {author} {\bibfnamefont {Z.}~\bibnamefont {Wang}},
  \bibinfo {author} {\bibfnamefont {M.~G.}\ \bibnamefont {Vergniory}}, \bibinfo
  {author} {\bibfnamefont {A.~M.}\ \bibnamefont {Cook}}, \bibinfo {author}
  {\bibfnamefont {A.}~\bibnamefont {Murani}}, \bibinfo {author} {\bibfnamefont
  {S.}~\bibnamefont {Sengupta}}, \bibinfo {author} {\bibfnamefont {A.~Y.}\
  \bibnamefont {Kasumov}}, \bibinfo {author} {\bibfnamefont {R.}~\bibnamefont
  {Deblock}}, \bibinfo {author} {\bibfnamefont {S.}~\bibnamefont {Jeon}},
  \bibinfo {author} {\bibfnamefont {I.}~\bibnamefont {Drozdov}}, \bibinfo
  {author} {\bibfnamefont {H.}~\bibnamefont {Bouchiat}}, \bibinfo {author}
  {\bibfnamefont {S.}~\bibnamefont {Gu{\'{e}}ron}}, \bibinfo {author}
  {\bibfnamefont {A.}~\bibnamefont {Yazdani}}, \bibinfo {author} {\bibfnamefont
  {B.~A.}\ \bibnamefont {Bernevig}}, \ and\ \bibinfo {author} {\bibfnamefont
  {T.}~\bibnamefont {Neupert}},\ }\href {\doibase 10.1038/s41567-018-0224-7}
  {\bibfield  {journal} {\bibinfo  {journal} {Nature Phys.}\ }\textbf {\bibinfo
  {volume} {14}},\ \bibinfo {pages} {918} (\bibinfo {year} {2018})},\ \Eprint
  {http://arxiv.org/abs/1802.02585} {arXiv:1802.02585} \BibitemShut {NoStop}%
\bibitem [{\citenamefont {Hsu}\ \emph {et~al.}(2019)\citenamefont {Hsu},
  \citenamefont {Zhou}, \citenamefont {Chang}, \citenamefont {Ma},
  \citenamefont {Gedik}, \citenamefont {Bansil}, \citenamefont {Xu},
  \citenamefont {Lin},\ and\ \citenamefont {Fu}}]{hsu2019topology}%
  \BibitemOpen
  \bibfield  {author} {\bibinfo {author} {\bibfnamefont {C.-H.}\ \bibnamefont
  {Hsu}}, \bibinfo {author} {\bibfnamefont {X.}~\bibnamefont {Zhou}}, \bibinfo
  {author} {\bibfnamefont {T.-R.}\ \bibnamefont {Chang}}, \bibinfo {author}
  {\bibfnamefont {Q.}~\bibnamefont {Ma}}, \bibinfo {author} {\bibfnamefont
  {N.}~\bibnamefont {Gedik}}, \bibinfo {author} {\bibfnamefont
  {A.}~\bibnamefont {Bansil}}, \bibinfo {author} {\bibfnamefont {S.-Y.}\
  \bibnamefont {Xu}}, \bibinfo {author} {\bibfnamefont {H.}~\bibnamefont
  {Lin}}, \ and\ \bibinfo {author} {\bibfnamefont {L.}~\bibnamefont {Fu}},\
  }\href@noop {} {\bibfield  {journal} {\bibinfo  {journal} {Proc. Natl. Acad.
  Sci. U.S.A.}\ }\textbf {\bibinfo {volume} {116}},\ \bibinfo {pages} {13255}
  (\bibinfo {year} {2019})}\BibitemShut {NoStop}%
\bibitem [{\citenamefont {Hirahara}\ \emph
  {et~al.}(2011{\natexlab{a}})\citenamefont {Hirahara}, \citenamefont
  {Bihlmayer}, \citenamefont {Sakamoto}, \citenamefont {Yamada}, \citenamefont
  {Miyazaki}, \citenamefont {Kimura}, \citenamefont {Bl{\"{u}}gel},\ and\
  \citenamefont {Hasegawa}}]{Hirahara2011a}%
  \BibitemOpen
  \bibfield  {author} {\bibinfo {author} {\bibfnamefont {T.}~\bibnamefont
  {Hirahara}}, \bibinfo {author} {\bibfnamefont {G.}~\bibnamefont {Bihlmayer}},
  \bibinfo {author} {\bibfnamefont {Y.}~\bibnamefont {Sakamoto}}, \bibinfo
  {author} {\bibfnamefont {M.}~\bibnamefont {Yamada}}, \bibinfo {author}
  {\bibfnamefont {H.}~\bibnamefont {Miyazaki}}, \bibinfo {author}
  {\bibfnamefont {S.-i.}\ \bibnamefont {Kimura}}, \bibinfo {author}
  {\bibfnamefont {S.}~\bibnamefont {Bl{\"{u}}gel}}, \ and\ \bibinfo {author}
  {\bibfnamefont {S.}~\bibnamefont {Hasegawa}},\ }\href {\doibase
  10.1103/PhysRevLett.107.166801} {\bibfield  {journal} {\bibinfo  {journal}
  {Phys. Rev. Lett.}\ }\textbf {\bibinfo {volume} {107}},\ \bibinfo {pages}
  {166801} (\bibinfo {year} {2011}{\natexlab{a}})}\BibitemShut {NoStop}%
\bibitem [{\citenamefont {Wang}\ \emph {et~al.}(2014)\citenamefont {Wang},
  \citenamefont {Chen},\ and\ \citenamefont {Liu}}]{Wang2014a}%
  \BibitemOpen
  \bibfield  {author} {\bibinfo {author} {\bibfnamefont {Z.~F.}\ \bibnamefont
  {Wang}}, \bibinfo {author} {\bibfnamefont {L.}~\bibnamefont {Chen}}, \ and\
  \bibinfo {author} {\bibfnamefont {F.}~\bibnamefont {Liu}},\ }\href {\doibase
  10.1021/nl5009212} {\bibfield  {journal} {\bibinfo  {journal} {Nano Lett.}\
  }\textbf {\bibinfo {volume} {14}},\ \bibinfo {pages} {2879} (\bibinfo {year}
  {2014})},\ \Eprint {http://arxiv.org/abs/1402.6754} {arXiv:1402.6754}
  \BibitemShut {NoStop}%
\bibitem [{\citenamefont {Ma}\ \emph {et~al.}(2015)\citenamefont {Ma},
  \citenamefont {Dai}, \citenamefont {Kou}, \citenamefont {Frauenheim},\ and\
  \citenamefont {Heine}}]{Ma2015a}%
  \BibitemOpen
  \bibfield  {author} {\bibinfo {author} {\bibfnamefont {Y.}~\bibnamefont
  {Ma}}, \bibinfo {author} {\bibfnamefont {Y.}~\bibnamefont {Dai}}, \bibinfo
  {author} {\bibfnamefont {L.}~\bibnamefont {Kou}}, \bibinfo {author}
  {\bibfnamefont {T.}~\bibnamefont {Frauenheim}}, \ and\ \bibinfo {author}
  {\bibfnamefont {T.}~\bibnamefont {Heine}},\ }\href {\doibase
  10.1021/nl504037u} {\bibfield  {journal} {\bibinfo  {journal} {Nano Lett.}\
  }\textbf {\bibinfo {volume} {15}},\ \bibinfo {pages} {1083} (\bibinfo {year}
  {2015})}\BibitemShut {NoStop}%
\bibitem [{\citenamefont {Li}\ \emph {et~al.}(2017)\citenamefont {Li},
  \citenamefont {Ji}, \citenamefont {Li}, \citenamefont {Hu}, \citenamefont
  {Cai}, \citenamefont {Zhang},\ and\ \citenamefont {Yan}}]{Li2017a}%
  \BibitemOpen
  \bibfield  {author} {\bibinfo {author} {\bibfnamefont {S.-s.}\ \bibnamefont
  {Li}}, \bibinfo {author} {\bibfnamefont {W.-x.}\ \bibnamefont {Ji}}, \bibinfo
  {author} {\bibfnamefont {P.}~\bibnamefont {Li}}, \bibinfo {author}
  {\bibfnamefont {S.-j.}\ \bibnamefont {Hu}}, \bibinfo {author} {\bibfnamefont
  {L.}~\bibnamefont {Cai}}, \bibinfo {author} {\bibfnamefont {C.-w.}\
  \bibnamefont {Zhang}}, \ and\ \bibinfo {author} {\bibfnamefont {S.-s.}\
  \bibnamefont {Yan}},\ }\href {\doibase 10.1021/acsami.7b02818} {\bibfield
  {journal} {\bibinfo  {journal} {ACS Appl. Mater. Interfaces}\ }\textbf
  {\bibinfo {volume} {9}},\ \bibinfo {pages} {21515} (\bibinfo {year}
  {2017})}\BibitemShut {NoStop}%
\bibitem [{\citenamefont {Lu}\ \emph {et~al.}(2015)\citenamefont {Lu},
  \citenamefont {Xu}, \citenamefont {Zeng}, \citenamefont {Yao}, \citenamefont
  {Shen}, \citenamefont {Yang}, \citenamefont {Luo}, \citenamefont {Pan},
  \citenamefont {Wu}, \citenamefont {Das}, \citenamefont {He}, \citenamefont
  {Jiang}, \citenamefont {Martin}, \citenamefont {Feng}, \citenamefont {Lin},\
  and\ \citenamefont {Wang}}]{Lu2015a}%
  \BibitemOpen
  \bibfield  {author} {\bibinfo {author} {\bibfnamefont {Y.}~\bibnamefont
  {Lu}}, \bibinfo {author} {\bibfnamefont {W.}~\bibnamefont {Xu}}, \bibinfo
  {author} {\bibfnamefont {M.}~\bibnamefont {Zeng}}, \bibinfo {author}
  {\bibfnamefont {G.}~\bibnamefont {Yao}}, \bibinfo {author} {\bibfnamefont
  {L.}~\bibnamefont {Shen}}, \bibinfo {author} {\bibfnamefont {M.}~\bibnamefont
  {Yang}}, \bibinfo {author} {\bibfnamefont {Z.}~\bibnamefont {Luo}}, \bibinfo
  {author} {\bibfnamefont {F.}~\bibnamefont {Pan}}, \bibinfo {author}
  {\bibfnamefont {K.}~\bibnamefont {Wu}}, \bibinfo {author} {\bibfnamefont
  {T.}~\bibnamefont {Das}}, \bibinfo {author} {\bibfnamefont {P.}~\bibnamefont
  {He}}, \bibinfo {author} {\bibfnamefont {J.}~\bibnamefont {Jiang}}, \bibinfo
  {author} {\bibfnamefont {J.}~\bibnamefont {Martin}}, \bibinfo {author}
  {\bibfnamefont {Y.~P.}\ \bibnamefont {Feng}}, \bibinfo {author}
  {\bibfnamefont {H.}~\bibnamefont {Lin}}, \ and\ \bibinfo {author}
  {\bibfnamefont {X.-s.}\ \bibnamefont {Wang}},\ }\href {\doibase
  10.1021/nl502997v} {\bibfield  {journal} {\bibinfo  {journal} {Nano Lett.}\
  }\textbf {\bibinfo {volume} {15}},\ \bibinfo {pages} {80} (\bibinfo {year}
  {2015})}\BibitemShut {NoStop}%
\bibitem [{\citenamefont {Saito}\ \emph {et~al.}(2016)\citenamefont {Saito},
  \citenamefont {Sawahata}, \citenamefont {Komine},\ and\ \citenamefont
  {Aono}}]{PhysRevB.93.041301}%
  \BibitemOpen
  \bibfield  {author} {\bibinfo {author} {\bibfnamefont {K.}~\bibnamefont
  {Saito}}, \bibinfo {author} {\bibfnamefont {H.}~\bibnamefont {Sawahata}},
  \bibinfo {author} {\bibfnamefont {T.}~\bibnamefont {Komine}}, \ and\ \bibinfo
  {author} {\bibfnamefont {T.}~\bibnamefont {Aono}},\ }\href {\doibase
  10.1103/PhysRevB.93.041301} {\bibfield  {journal} {\bibinfo  {journal} {Phys.
  Rev. B}\ }\textbf {\bibinfo {volume} {93}},\ \bibinfo {pages} {041301}
  (\bibinfo {year} {2016})}\BibitemShut {NoStop}%
\bibitem [{\citenamefont {Nouri}\ \emph {et~al.}(2020)\citenamefont {Nouri},
  \citenamefont {Rashedi},\ and\ \citenamefont {Karbaschi}}]{NOURI2020126364}%
  \BibitemOpen
  \bibfield  {author} {\bibinfo {author} {\bibfnamefont {N.}~\bibnamefont
  {Nouri}}, \bibinfo {author} {\bibfnamefont {G.}~\bibnamefont {Rashedi}}, \
  and\ \bibinfo {author} {\bibfnamefont {H.}~\bibnamefont {Karbaschi}},\ }\href
  {\doibase https://doi.org/10.1016/j.physleta.2020.126364} {\bibfield
  {journal} {\bibinfo  {journal} {Phys. Lett. A}\ }\textbf {\bibinfo {volume}
  {384}},\ \bibinfo {pages} {126364} (\bibinfo {year} {2020})}\BibitemShut
  {NoStop}%
\bibitem [{\citenamefont {Bieniek}\ \emph {et~al.}(2017)\citenamefont
  {Bieniek}, \citenamefont {Wo{\'{z}}niak},\ and\ \citenamefont
  {Potasz}}]{Bieniek2017a}%
  \BibitemOpen
  \bibfield  {author} {\bibinfo {author} {\bibfnamefont {M.}~\bibnamefont
  {Bieniek}}, \bibinfo {author} {\bibfnamefont {T.}~\bibnamefont
  {Wo{\'{z}}niak}}, \ and\ \bibinfo {author} {\bibfnamefont {P.}~\bibnamefont
  {Potasz}},\ }\href {\doibase 10.1088/1361-648X/aa5e79} {\bibfield  {journal}
  {\bibinfo  {journal} {J. Phys. Condens. Matter}\ }\textbf {\bibinfo {volume}
  {29}},\ \bibinfo {pages} {155501} (\bibinfo {year} {2017})},\ \Eprint
  {http://arxiv.org/abs/1612.00647} {arXiv:1612.00647} \BibitemShut {NoStop}%
\bibitem [{\citenamefont {Mostofi}\ \emph {et~al.}(2008)\citenamefont
  {Mostofi}, \citenamefont {Yates}, \citenamefont {Lee}, \citenamefont {Souza},
  \citenamefont {Vanderbilt},\ and\ \citenamefont {Marzari}}]{Mostofi2008a}%
  \BibitemOpen
  \bibfield  {author} {\bibinfo {author} {\bibfnamefont {A.~A.}\ \bibnamefont
  {Mostofi}}, \bibinfo {author} {\bibfnamefont {J.~R.}\ \bibnamefont {Yates}},
  \bibinfo {author} {\bibfnamefont {Y.-S.}\ \bibnamefont {Lee}}, \bibinfo
  {author} {\bibfnamefont {I.}~\bibnamefont {Souza}}, \bibinfo {author}
  {\bibfnamefont {D.}~\bibnamefont {Vanderbilt}}, \ and\ \bibinfo {author}
  {\bibfnamefont {N.}~\bibnamefont {Marzari}},\ }\href {\doibase
  10.1016/j.cpc.2007.11.016} {\bibfield  {journal} {\bibinfo  {journal}
  {Comput. Phys. Commun.}\ }\textbf {\bibinfo {volume} {178}},\ \bibinfo
  {pages} {685} (\bibinfo {year} {2008})}\BibitemShut {NoStop}%
\bibitem [{\citenamefont {Zak}(1989)}]{Zak1989a}%
  \BibitemOpen
  \bibfield  {author} {\bibinfo {author} {\bibfnamefont {J.}~\bibnamefont
  {Zak}},\ }\href {\doibase 10.1103/PhysRevLett.62.2747} {\bibfield  {journal}
  {\bibinfo  {journal} {Phys. Rev. Lett.}\ }\textbf {\bibinfo {volume} {62}},\
  \bibinfo {pages} {2747} (\bibinfo {year} {1989})}\BibitemShut {NoStop}%
\bibitem [{\citenamefont {Klymenko}\ \emph {et~al.}(2021)\citenamefont
  {Klymenko}, \citenamefont {Vaitkus}, \citenamefont {Smith},\ and\
  \citenamefont {Cole}}]{KLYMENKO2021107676}%
  \BibitemOpen
  \bibfield  {author} {\bibinfo {author} {\bibfnamefont {M.}~\bibnamefont
  {Klymenko}}, \bibinfo {author} {\bibfnamefont {J.}~\bibnamefont {Vaitkus}},
  \bibinfo {author} {\bibfnamefont {J.}~\bibnamefont {Smith}}, \ and\ \bibinfo
  {author} {\bibfnamefont {J.}~\bibnamefont {Cole}},\ }\href {\doibase
  https://doi.org/10.1016/j.cpc.2020.107676} {\bibfield  {journal} {\bibinfo
  {journal} {Comput. Phys. Commun.}\ }\textbf {\bibinfo {volume} {259}},\
  \bibinfo {pages} {107676} (\bibinfo {year} {2021})}\BibitemShut {NoStop}%
\bibitem [{\citenamefont {Wada}\ \emph {et~al.}(2011)\citenamefont {Wada},
  \citenamefont {Murakami}, \citenamefont {Freimuth},\ and\ \citenamefont
  {Bihlmayer}}]{PhysRevB.83.121310}%
  \BibitemOpen
  \bibfield  {author} {\bibinfo {author} {\bibfnamefont {M.}~\bibnamefont
  {Wada}}, \bibinfo {author} {\bibfnamefont {S.}~\bibnamefont {Murakami}},
  \bibinfo {author} {\bibfnamefont {F.}~\bibnamefont {Freimuth}}, \ and\
  \bibinfo {author} {\bibfnamefont {G.}~\bibnamefont {Bihlmayer}},\ }\href
  {\doibase 10.1103/PhysRevB.83.121310} {\bibfield  {journal} {\bibinfo
  {journal} {Phys. Rev. B}\ }\textbf {\bibinfo {volume} {83}},\ \bibinfo
  {pages} {121310} (\bibinfo {year} {2011})}\BibitemShut {NoStop}%
\bibitem [{\citenamefont {Li}\ \emph {et~al.}(2014)\citenamefont {Li},
  \citenamefont {Liu}, \citenamefont {Jiang}, \citenamefont {Wang},\ and\
  \citenamefont {Feng}}]{Li2014a}%
  \BibitemOpen
  \bibfield  {author} {\bibinfo {author} {\bibfnamefont {X.}~\bibnamefont
  {Li}}, \bibinfo {author} {\bibfnamefont {H.}~\bibnamefont {Liu}}, \bibinfo
  {author} {\bibfnamefont {H.}~\bibnamefont {Jiang}}, \bibinfo {author}
  {\bibfnamefont {F.}~\bibnamefont {Wang}}, \ and\ \bibinfo {author}
  {\bibfnamefont {J.}~\bibnamefont {Feng}},\ }\href {\doibase
  10.1103/PhysRevB.90.165412} {\bibfield  {journal} {\bibinfo  {journal} {Phys.
  Rev. B}\ }\textbf {\bibinfo {volume} {90}},\ \bibinfo {pages} {165412}
  (\bibinfo {year} {2014})}\BibitemShut {NoStop}%
\bibitem [{\citenamefont {Hirahara}\ \emph
  {et~al.}(2011{\natexlab{b}})\citenamefont {Hirahara}, \citenamefont
  {Bihlmayer}, \citenamefont {Sakamoto}, \citenamefont {Yamada}, \citenamefont
  {Miyazaki}, \citenamefont {Kimura}, \citenamefont {Bl\"ugel},\ and\
  \citenamefont {Hasegawa}}]{PhysRevLett.107.166801}%
  \BibitemOpen
  \bibfield  {author} {\bibinfo {author} {\bibfnamefont {T.}~\bibnamefont
  {Hirahara}}, \bibinfo {author} {\bibfnamefont {G.}~\bibnamefont {Bihlmayer}},
  \bibinfo {author} {\bibfnamefont {Y.}~\bibnamefont {Sakamoto}}, \bibinfo
  {author} {\bibfnamefont {M.}~\bibnamefont {Yamada}}, \bibinfo {author}
  {\bibfnamefont {H.}~\bibnamefont {Miyazaki}}, \bibinfo {author}
  {\bibfnamefont {S.-i.}\ \bibnamefont {Kimura}}, \bibinfo {author}
  {\bibfnamefont {S.}~\bibnamefont {Bl\"ugel}}, \ and\ \bibinfo {author}
  {\bibfnamefont {S.}~\bibnamefont {Hasegawa}},\ }\href {\doibase
  10.1103/PhysRevLett.107.166801} {\bibfield  {journal} {\bibinfo  {journal}
  {Phys. Rev. Lett.}\ }\textbf {\bibinfo {volume} {107}},\ \bibinfo {pages}
  {166801} (\bibinfo {year} {2011}{\natexlab{b}})}\BibitemShut {NoStop}%
\bibitem [{\citenamefont {Remediakis}\ and\ \citenamefont
  {Kaxiras}(1999)}]{Remediakis1999a}%
  \BibitemOpen
  \bibfield  {author} {\bibinfo {author} {\bibfnamefont {I.~N.}\ \bibnamefont
  {Remediakis}}\ and\ \bibinfo {author} {\bibfnamefont {E.}~\bibnamefont
  {Kaxiras}},\ }\href {\doibase 10.1103/PhysRevB.59.5536} {\bibfield  {journal}
  {\bibinfo  {journal} {Phys. Rev. B}\ }\textbf {\bibinfo {volume} {59}},\
  \bibinfo {pages} {5536} (\bibinfo {year} {1999})}\BibitemShut {NoStop}%
\bibitem [{\citenamefont {Yang}\ \emph {et~al.}(2012)\citenamefont {Yang},
  \citenamefont {Miao}, \citenamefont {Wang}, \citenamefont {Yao},
  \citenamefont {Zhu}, \citenamefont {Song}, \citenamefont {Wang},
  \citenamefont {Xu}, \citenamefont {Fedorov}, \citenamefont {Sun},
  \citenamefont {Zhang}, \citenamefont {Liu}, \citenamefont {Liu},
  \citenamefont {Qian}, \citenamefont {Gao},\ and\ \citenamefont
  {Jia}}]{PhysRevLett.109.016801}%
  \BibitemOpen
  \bibfield  {author} {\bibinfo {author} {\bibfnamefont {F.}~\bibnamefont
  {Yang}}, \bibinfo {author} {\bibfnamefont {L.}~\bibnamefont {Miao}}, \bibinfo
  {author} {\bibfnamefont {Z.~F.}\ \bibnamefont {Wang}}, \bibinfo {author}
  {\bibfnamefont {M.-Y.}\ \bibnamefont {Yao}}, \bibinfo {author} {\bibfnamefont
  {F.}~\bibnamefont {Zhu}}, \bibinfo {author} {\bibfnamefont {Y.~R.}\
  \bibnamefont {Song}}, \bibinfo {author} {\bibfnamefont {M.-X.}\ \bibnamefont
  {Wang}}, \bibinfo {author} {\bibfnamefont {J.-P.}\ \bibnamefont {Xu}},
  \bibinfo {author} {\bibfnamefont {A.~V.}\ \bibnamefont {Fedorov}}, \bibinfo
  {author} {\bibfnamefont {Z.}~\bibnamefont {Sun}}, \bibinfo {author}
  {\bibfnamefont {G.~B.}\ \bibnamefont {Zhang}}, \bibinfo {author}
  {\bibfnamefont {C.}~\bibnamefont {Liu}}, \bibinfo {author} {\bibfnamefont
  {F.}~\bibnamefont {Liu}}, \bibinfo {author} {\bibfnamefont {D.}~\bibnamefont
  {Qian}}, \bibinfo {author} {\bibfnamefont {C.~L.}\ \bibnamefont {Gao}}, \
  and\ \bibinfo {author} {\bibfnamefont {J.-F.}\ \bibnamefont {Jia}},\ }\href
  {\doibase 10.1103/PhysRevLett.109.016801} {\bibfield  {journal} {\bibinfo
  {journal} {Phys. Rev. Lett.}\ }\textbf {\bibinfo {volume} {109}},\ \bibinfo
  {pages} {016801} (\bibinfo {year} {2012})}\BibitemShut {NoStop}%
\bibitem [{\citenamefont {Kresse}\ and\ \citenamefont
  {Furthm{\"{u}}ller}(1996{\natexlab{a}})}]{Kresse1996a}%
  \BibitemOpen
  \bibfield  {author} {\bibinfo {author} {\bibfnamefont {G.}~\bibnamefont
  {Kresse}}\ and\ \bibinfo {author} {\bibfnamefont {J.}~\bibnamefont
  {Furthm{\"{u}}ller}},\ }\href {\doibase 10.1103/PhysRevB.54.11169} {\bibfield
   {journal} {\bibinfo  {journal} {Phys. Rev. B}\ }\textbf {\bibinfo {volume}
  {54}},\ \bibinfo {pages} {11169} (\bibinfo {year}
  {1996}{\natexlab{a}})}\BibitemShut {NoStop}%
\bibitem [{\citenamefont {Kresse}\ and\ \citenamefont
  {Furthm{\"{u}}ller}(1996{\natexlab{b}})}]{Kresse1996b}%
  \BibitemOpen
  \bibfield  {author} {\bibinfo {author} {\bibfnamefont {G.}~\bibnamefont
  {Kresse}}\ and\ \bibinfo {author} {\bibfnamefont {J.}~\bibnamefont
  {Furthm{\"{u}}ller}},\ }\href {\doibase 10.1016/0927-0256(96)00008-0}
  {\bibfield  {journal} {\bibinfo  {journal} {Comput. Mater. Sci.}\ }\textbf
  {\bibinfo {volume} {6}},\ \bibinfo {pages} {15} (\bibinfo {year}
  {1996}{\natexlab{b}})}\BibitemShut {NoStop}%
\bibitem [{\citenamefont {Perdew}\ \emph {et~al.}(1996)\citenamefont {Perdew},
  \citenamefont {Burke},\ and\ \citenamefont {Ernzerhof}}]{Perdew1996a}%
  \BibitemOpen
  \bibfield  {author} {\bibinfo {author} {\bibfnamefont {J.~P.}\ \bibnamefont
  {Perdew}}, \bibinfo {author} {\bibfnamefont {K.}~\bibnamefont {Burke}}, \
  and\ \bibinfo {author} {\bibfnamefont {M.}~\bibnamefont {Ernzerhof}},\ }\href
  {\doibase 10.1103/PhysRevLett.77.3865} {\bibfield  {journal} {\bibinfo
  {journal} {Phys. Rev. Lett.}\ }\textbf {\bibinfo {volume} {77}},\ \bibinfo
  {pages} {3865} (\bibinfo {year} {1996})}\BibitemShut {NoStop}%
\bibitem [{\citenamefont {Mostofi}\ \emph {et~al.}(2014)\citenamefont
  {Mostofi}, \citenamefont {Yates}, \citenamefont {Pizzi}, \citenamefont {Lee},
  \citenamefont {Souza}, \citenamefont {Vanderbilt},\ and\ \citenamefont
  {Marzari}}]{Mostofi2014a}%
  \BibitemOpen
  \bibfield  {author} {\bibinfo {author} {\bibfnamefont {A.~A.}\ \bibnamefont
  {Mostofi}}, \bibinfo {author} {\bibfnamefont {J.~R.}\ \bibnamefont {Yates}},
  \bibinfo {author} {\bibfnamefont {G.}~\bibnamefont {Pizzi}}, \bibinfo
  {author} {\bibfnamefont {Y.-S.}\ \bibnamefont {Lee}}, \bibinfo {author}
  {\bibfnamefont {I.}~\bibnamefont {Souza}}, \bibinfo {author} {\bibfnamefont
  {D.}~\bibnamefont {Vanderbilt}}, \ and\ \bibinfo {author} {\bibfnamefont
  {N.}~\bibnamefont {Marzari}},\ }\href {\doibase 10.1016/j.cpc.2014.05.003}
  {\bibfield  {journal} {\bibinfo  {journal} {Comput. Phys. Commun.}\ }\textbf
  {\bibinfo {volume} {185}},\ \bibinfo {pages} {2309} (\bibinfo {year}
  {2014})}\BibitemShut {NoStop}%
\bibitem [{\citenamefont {Marzari}\ \emph {et~al.}(2012)\citenamefont
  {Marzari}, \citenamefont {Mostofi}, \citenamefont {Yates}, \citenamefont
  {Souza},\ and\ \citenamefont {Vanderbilt}}]{Marzari2012a}%
  \BibitemOpen
  \bibfield  {author} {\bibinfo {author} {\bibfnamefont {N.}~\bibnamefont
  {Marzari}}, \bibinfo {author} {\bibfnamefont {A.~A.}\ \bibnamefont
  {Mostofi}}, \bibinfo {author} {\bibfnamefont {J.~R.}\ \bibnamefont {Yates}},
  \bibinfo {author} {\bibfnamefont {I.}~\bibnamefont {Souza}}, \ and\ \bibinfo
  {author} {\bibfnamefont {D.}~\bibnamefont {Vanderbilt}},\ }\href {\doibase
  10.1103/RevModPhys.84.1419} {\bibfield  {journal} {\bibinfo  {journal}
  {Reviews of Modern Physics}\ }\textbf {\bibinfo {volume} {84}},\ \bibinfo
  {pages} {1419} (\bibinfo {year} {2012})}\BibitemShut {NoStop}%
\bibitem [{\citenamefont {Gresch}\ \emph {et~al.}(2018)\citenamefont {Gresch},
  \citenamefont {Wu}, \citenamefont {Winkler}, \citenamefont
  {H{\"{a}}uselmann}, \citenamefont {Troyer},\ and\ \citenamefont
  {Soluyanov}}]{Gresch2018a}%
  \BibitemOpen
  \bibfield  {author} {\bibinfo {author} {\bibfnamefont {D.}~\bibnamefont
  {Gresch}}, \bibinfo {author} {\bibfnamefont {Q.}~\bibnamefont {Wu}}, \bibinfo
  {author} {\bibfnamefont {G.~W.}\ \bibnamefont {Winkler}}, \bibinfo {author}
  {\bibfnamefont {R.}~\bibnamefont {H{\"{a}}uselmann}}, \bibinfo {author}
  {\bibfnamefont {M.}~\bibnamefont {Troyer}}, \ and\ \bibinfo {author}
  {\bibfnamefont {A.~A.}\ \bibnamefont {Soluyanov}},\ }\href {\doibase
  10.1103/PhysRevMaterials.2.103805} {\bibfield  {journal} {\bibinfo  {journal}
  {Phys. Rev. Mater.}\ }\textbf {\bibinfo {volume} {2}},\ \bibinfo {pages}
  {103805} (\bibinfo {year} {2018})}\BibitemShut {NoStop}%
\bibitem [{\citenamefont {Dresselhaus}\ \emph {et~al.}(2008)\citenamefont
  {Dresselhaus}, \citenamefont {Dresselhaus},\ and\ \citenamefont
  {Jorio}}]{Dresselhaus2008a}%
  \BibitemOpen
  \bibfield  {author} {\bibinfo {author} {\bibfnamefont {M.~S.}\ \bibnamefont
  {Dresselhaus}}, \bibinfo {author} {\bibfnamefont {G.}~\bibnamefont
  {Dresselhaus}}, \ and\ \bibinfo {author} {\bibfnamefont {A.}~\bibnamefont
  {Jorio}},\ }\href {\doibase 10.1007/978-3-540-32899-8} {\emph {\bibinfo
  {title} {{Group Theroy: Application to the Physics of Condensed Matter}}}},\
  \bibinfo {edition} {1st}\ ed.\ (\bibinfo  {publisher} {Springer-Verlag},\
  \bibinfo {address} {Berlin Heidelberg},\ \bibinfo {year} {2008})\BibitemShut
  {NoStop}%
\bibitem [{\citenamefont {Kochan}\ \emph {et~al.}(2017)\citenamefont {Kochan},
  \citenamefont {Irmer},\ and\ \citenamefont {Fabian}}]{soc_pg}%
  \BibitemOpen
  \bibfield  {author} {\bibinfo {author} {\bibfnamefont {D.}~\bibnamefont
  {Kochan}}, \bibinfo {author} {\bibfnamefont {S.}~\bibnamefont {Irmer}}, \
  and\ \bibinfo {author} {\bibfnamefont {J.}~\bibnamefont {Fabian}},\ }\href
  {\doibase 10.1103/PhysRevB.95.165415} {\bibfield  {journal} {\bibinfo
  {journal} {Phys. Rev. B}\ }\textbf {\bibinfo {volume} {95}},\ \bibinfo
  {pages} {165415} (\bibinfo {year} {2017})}\BibitemShut {NoStop}%
\bibitem [{\citenamefont {Koster;}(1963)}]{koster}%
  \BibitemOpen
  \bibfield  {author} {\bibinfo {author} {\bibfnamefont {G.~F.}\ \bibnamefont
  {Koster;}},\ }\href@noop {} {\emph {\bibinfo {title} {Properties of the
  Thirty-two Point Groups}}}\ (\bibinfo  {publisher} {The M.I.T. Press},\
  \bibinfo {year} {1963})\BibitemShut {NoStop}%
\bibitem [{\citenamefont {Fu}\ and\ \citenamefont {Kane}(2007)}]{Fu2007b}%
  \BibitemOpen
  \bibfield  {author} {\bibinfo {author} {\bibfnamefont {L.}~\bibnamefont
  {Fu}}\ and\ \bibinfo {author} {\bibfnamefont {C.~L.}\ \bibnamefont {Kane}},\
  }\href {\doibase 10.1103/PhysRevB.76.045302} {\bibfield  {journal} {\bibinfo
  {journal} {Phys. Rev. B}\ }\textbf {\bibinfo {volume} {76}},\ \bibinfo
  {pages} {045302} (\bibinfo {year} {2007})}\BibitemShut {NoStop}%
\bibitem [{\citenamefont {Huang}\ \emph {et~al.}(2013)\citenamefont {Huang},
  \citenamefont {Chuang}, \citenamefont {Hsu}, \citenamefont {Liu},
  \citenamefont {Chang}, \citenamefont {Lin},\ and\ \citenamefont
  {Bansil}}]{Huang2013a}%
  \BibitemOpen
  \bibfield  {author} {\bibinfo {author} {\bibfnamefont {Z.-Q.}\ \bibnamefont
  {Huang}}, \bibinfo {author} {\bibfnamefont {F.-C.}\ \bibnamefont {Chuang}},
  \bibinfo {author} {\bibfnamefont {C.-H.}\ \bibnamefont {Hsu}}, \bibinfo
  {author} {\bibfnamefont {Y.-T.}\ \bibnamefont {Liu}}, \bibinfo {author}
  {\bibfnamefont {H.-R.}\ \bibnamefont {Chang}}, \bibinfo {author}
  {\bibfnamefont {H.}~\bibnamefont {Lin}}, \ and\ \bibinfo {author}
  {\bibfnamefont {A.}~\bibnamefont {Bansil}},\ }\href {\doibase
  10.1103/PhysRevB.88.165301} {\bibfield  {journal} {\bibinfo  {journal} {Phys.
  Rev. B}\ }\textbf {\bibinfo {volume} {88}},\ \bibinfo {pages} {165301}
  (\bibinfo {year} {2013})}\BibitemShut {NoStop}%
\bibitem [{\citenamefont {Kowalczyk}\ \emph {et~al.}(2020)\citenamefont
  {Kowalczyk}, \citenamefont {Brown}, \citenamefont {Maerkl}, \citenamefont
  {Lu}, \citenamefont {Chiu}, \citenamefont {Liu}, \citenamefont {Yang},
  \citenamefont {Wang}, \citenamefont {Zasada}, \citenamefont {Genuzio},
  \citenamefont {Mente{\c s}}, \citenamefont {Locatelli}, \citenamefont
  {Chiang},\ and\ \citenamefont {Bian}}]{exp110}%
  \BibitemOpen
  \bibfield  {author} {\bibinfo {author} {\bibfnamefont {P.~J.}\ \bibnamefont
  {Kowalczyk}}, \bibinfo {author} {\bibfnamefont {S.~A.}\ \bibnamefont
  {Brown}}, \bibinfo {author} {\bibfnamefont {T.}~\bibnamefont {Maerkl}},
  \bibinfo {author} {\bibfnamefont {Q.}~\bibnamefont {Lu}}, \bibinfo {author}
  {\bibfnamefont {C.-K.}\ \bibnamefont {Chiu}}, \bibinfo {author}
  {\bibfnamefont {Y.}~\bibnamefont {Liu}}, \bibinfo {author} {\bibfnamefont
  {S.~A.}\ \bibnamefont {Yang}}, \bibinfo {author} {\bibfnamefont
  {X.}~\bibnamefont {Wang}}, \bibinfo {author} {\bibfnamefont {I.}~\bibnamefont
  {Zasada}}, \bibinfo {author} {\bibfnamefont {F.}~\bibnamefont {Genuzio}},
  \bibinfo {author} {\bibfnamefont {T.~O.}\ \bibnamefont {Mente{\c s}}},
  \bibinfo {author} {\bibfnamefont {A.}~\bibnamefont {Locatelli}}, \bibinfo
  {author} {\bibfnamefont {T.-C.}\ \bibnamefont {Chiang}}, \ and\ \bibinfo
  {author} {\bibfnamefont {G.}~\bibnamefont {Bian}},\ }\href {\doibase
  10.1021/acsnano.9b08136} {\bibfield  {journal} {\bibinfo  {journal} {ACS
  Nano}\ }\textbf {\bibinfo {volume} {14}},\ \bibinfo {pages} {1888} (\bibinfo
  {year} {2020})},\ \bibinfo {note} {pMID: 31971774}\BibitemShut {NoStop}%
\bibitem [{\citenamefont {Zhu}\ \emph {et~al.}(2019)\citenamefont {Zhu},
  \citenamefont {Shao}, \citenamefont {Wang}, \citenamefont {Cao},
  \citenamefont {Li}, \citenamefont {Liu}, \citenamefont {Liu}, \citenamefont
  {Liu}, \citenamefont {Wang}, \citenamefont {Ibrahim}, \citenamefont {Sun},
  \citenamefont {Wang}, \citenamefont {Du},\ and\ \citenamefont
  {Gao}}]{doi:10.1021/acs.nanolett.9b02444}%
  \BibitemOpen
  \bibfield  {author} {\bibinfo {author} {\bibfnamefont {S.-Y.}\ \bibnamefont
  {Zhu}}, \bibinfo {author} {\bibfnamefont {Y.}~\bibnamefont {Shao}}, \bibinfo
  {author} {\bibfnamefont {E.}~\bibnamefont {Wang}}, \bibinfo {author}
  {\bibfnamefont {L.}~\bibnamefont {Cao}}, \bibinfo {author} {\bibfnamefont
  {X.-Y.}\ \bibnamefont {Li}}, \bibinfo {author} {\bibfnamefont {Z.-L.}\
  \bibnamefont {Liu}}, \bibinfo {author} {\bibfnamefont {C.}~\bibnamefont
  {Liu}}, \bibinfo {author} {\bibfnamefont {L.-W.}\ \bibnamefont {Liu}},
  \bibinfo {author} {\bibfnamefont {J.-O.}\ \bibnamefont {Wang}}, \bibinfo
  {author} {\bibfnamefont {K.}~\bibnamefont {Ibrahim}}, \bibinfo {author}
  {\bibfnamefont {J.-T.}\ \bibnamefont {Sun}}, \bibinfo {author} {\bibfnamefont
  {Y.-L.}\ \bibnamefont {Wang}}, \bibinfo {author} {\bibfnamefont
  {S.}~\bibnamefont {Du}}, \ and\ \bibinfo {author} {\bibfnamefont {H.-J.}\
  \bibnamefont {Gao}},\ }\href {\doibase 10.1021/acs.nanolett.9b02444}
  {\bibfield  {journal} {\bibinfo  {journal} {Nano Lett.}\ }\textbf {\bibinfo
  {volume} {19}},\ \bibinfo {pages} {6323} (\bibinfo {year} {2019})},\ \bibinfo
  {note} {pMID: 31431010}\BibitemShut {NoStop}%
\bibitem [{\citenamefont {Zhang}\ \emph {et~al.}(2015)\citenamefont {Zhang},
  \citenamefont {Ma},\ and\ \citenamefont {Chen}}]{C5NR05006E}%
  \BibitemOpen
  \bibfield  {author} {\bibinfo {author} {\bibfnamefont {H.}~\bibnamefont
  {Zhang}}, \bibinfo {author} {\bibfnamefont {Y.}~\bibnamefont {Ma}}, \ and\
  \bibinfo {author} {\bibfnamefont {Z.}~\bibnamefont {Chen}},\ }\href {\doibase
  10.1039/C5NR05006E} {\bibfield  {journal} {\bibinfo  {journal} {Nanoscale}\
  }\textbf {\bibinfo {volume} {7}},\ \bibinfo {pages} {19152} (\bibinfo {year}
  {2015})}\BibitemShut {NoStop}%
\bibitem [{\citenamefont {Kou}\ \emph {et~al.}(2017)\citenamefont {Kou},
  \citenamefont {Ma}, \citenamefont {Sun}, \citenamefont {Heine},\ and\
  \citenamefont {Chen}}]{doi:10.1021/acs.jpclett.7b00222}%
  \BibitemOpen
  \bibfield  {author} {\bibinfo {author} {\bibfnamefont {L.}~\bibnamefont
  {Kou}}, \bibinfo {author} {\bibfnamefont {Y.}~\bibnamefont {Ma}}, \bibinfo
  {author} {\bibfnamefont {Z.}~\bibnamefont {Sun}}, \bibinfo {author}
  {\bibfnamefont {T.}~\bibnamefont {Heine}}, \ and\ \bibinfo {author}
  {\bibfnamefont {C.}~\bibnamefont {Chen}},\ }\href {\doibase
  10.1021/acs.jpclett.7b00222} {\bibfield  {journal} {\bibinfo  {journal} {J.
  Phys. Chem. Lett.}\ }\textbf {\bibinfo {volume} {8}},\ \bibinfo {pages}
  {1905} (\bibinfo {year} {2017})}\BibitemShut {NoStop}%
\bibitem [{\citenamefont {Acosta}\ and\ \citenamefont
  {Fazzio}(2019)}]{PhysRevLett.122.036401}%
  \BibitemOpen
  \bibfield  {author} {\bibinfo {author} {\bibfnamefont {C.~M.}\ \bibnamefont
  {Acosta}}\ and\ \bibinfo {author} {\bibfnamefont {A.}~\bibnamefont
  {Fazzio}},\ }\href {\doibase 10.1103/PhysRevLett.122.036401} {\bibfield
  {journal} {\bibinfo  {journal} {Phys. Rev. Lett.}\ }\textbf {\bibinfo
  {volume} {122}},\ \bibinfo {pages} {036401} (\bibinfo {year}
  {2019})}\BibitemShut {NoStop}%
\bibitem [{\citenamefont {Su}\ \emph {et~al.}(1979)\citenamefont {Su},
  \citenamefont {Schrieffer},\ and\ \citenamefont
  {Heeger}}]{PhysRevLett.42.1698}%
  \BibitemOpen
  \bibfield  {author} {\bibinfo {author} {\bibfnamefont {W.~P.}\ \bibnamefont
  {Su}}, \bibinfo {author} {\bibfnamefont {J.~R.}\ \bibnamefont {Schrieffer}},
  \ and\ \bibinfo {author} {\bibfnamefont {A.~J.}\ \bibnamefont {Heeger}},\
  }\href {\doibase 10.1103/PhysRevLett.42.1698} {\bibfield  {journal} {\bibinfo
   {journal} {Phys. Rev. Lett.}\ }\textbf {\bibinfo {volume} {42}},\ \bibinfo
  {pages} {1698} (\bibinfo {year} {1979})}\BibitemShut {NoStop}%
\bibitem [{\citenamefont {Gr{\"o}ning}\ \emph {et~al.}(2018)\citenamefont
  {Gr{\"o}ning}, \citenamefont {Wang}, \citenamefont {Yao}, \citenamefont
  {Pignedoli}, \citenamefont {Barin}, \citenamefont {Daniels}, \citenamefont
  {Cupo}, \citenamefont {Meunier}, \citenamefont {Feng}, \citenamefont {Narita}
  \emph {et~al.}}]{groning2018engineering}%
  \BibitemOpen
  \bibfield  {author} {\bibinfo {author} {\bibfnamefont {O.}~\bibnamefont
  {Gr{\"o}ning}}, \bibinfo {author} {\bibfnamefont {S.}~\bibnamefont {Wang}},
  \bibinfo {author} {\bibfnamefont {X.}~\bibnamefont {Yao}}, \bibinfo {author}
  {\bibfnamefont {C.~A.}\ \bibnamefont {Pignedoli}}, \bibinfo {author}
  {\bibfnamefont {G.~B.}\ \bibnamefont {Barin}}, \bibinfo {author}
  {\bibfnamefont {C.}~\bibnamefont {Daniels}}, \bibinfo {author} {\bibfnamefont
  {A.}~\bibnamefont {Cupo}}, \bibinfo {author} {\bibfnamefont {V.}~\bibnamefont
  {Meunier}}, \bibinfo {author} {\bibfnamefont {X.}~\bibnamefont {Feng}},
  \bibinfo {author} {\bibfnamefont {A.}~\bibnamefont {Narita}},  \emph
  {et~al.},\ }\href@noop {} {\bibfield  {journal} {\bibinfo  {journal}
  {Nature}\ }\textbf {\bibinfo {volume} {560}},\ \bibinfo {pages} {209}
  (\bibinfo {year} {2018})}\BibitemShut {NoStop}%
\bibitem [{\citenamefont {Zangeneh-Nejad}\ and\ \citenamefont
  {Fleury}(2020)}]{zangeneh2020disorder}%
  \BibitemOpen
  \bibfield  {author} {\bibinfo {author} {\bibfnamefont {F.}~\bibnamefont
  {Zangeneh-Nejad}}\ and\ \bibinfo {author} {\bibfnamefont {R.}~\bibnamefont
  {Fleury}},\ }\href@noop {} {\bibfield  {journal} {\bibinfo  {journal} {Adv.
  Mater.}\ }\textbf {\bibinfo {volume} {32}},\ \bibinfo {pages} {2001034}
  (\bibinfo {year} {2020})}\BibitemShut {NoStop}%
\end{thebibliography}
\end{document}